\documentclass[10pt,journal,letterpaper,compsoc]{IEEEtran}

\ifCLASSINFOpdf
\else
\fi

\newcommand{\eg}{{\it e.g., }}
\newcommand{\etal}{{\it et~al.}}
\newcommand{\ie}{{\it i.e., }}

\usepackage{balance}
\usepackage{times}
\usepackage{cite} 
\usepackage{pifont}
\usepackage{color, soul}
\usepackage{epsfig}
\usepackage{float}
\usepackage{graphicx} 
\usepackage{amsmath}
\usepackage{url}
\usepackage{subfigure}
\usepackage{amsthm}
\usepackage{algorithmicx}
\usepackage{algpseudocode}
\usepackage{algorithm}
\usepackage{xspace}
\usepackage{comment}
\floatstyle{boxed} 
\newfloat{textfig}{thp}{lop}
\newfloat{textfig}{thp}{lop}
\floatname{textfig}{\textbf{Figure}}
\usepackage{tikz}
\usepackage[]{hyperref}
\usepackage{mathtools}  
\usepackage{amssymb}
\usepackage{tabulary}
\usepackage{booktabs}

\newcommand\tikzmark[1]{%
  \tikz[remember picture,overlay]\node[inner sep=2pt] (#1) {};}

\algnewcommand\algorithmicinput{\textbf{Input:}}
\algnewcommand\INPUT{\item[\algorithmicinput]}

\algnewcommand\algorithmicoutput{\textbf{Output:}}
\algnewcommand\OUTPUT{\item[\algorithmicoutput]}

\newlength{\boxfigwidth}

\algdef{SE}[SUBALG]{Indent}{EndIndent}{}{\algorithmicend\ }%
\algtext*{Indent}
\algtext*{EndIndent}

\newcommand{\name}{CVS2\xspace}
\begin{document}
\title{Cost-Efficient and Robust On-Demand Video Stream Transcoding Using Heterogeneous Cloud Services}

\author{Xiangbo Li,
            Mohsen Amini Salehi,~\IEEEmembership{Member,~IEEE,}  
            Magdy Bayoumi,~\IEEEmembership{Fellow,~IEEE,} \\
            Nian-Feng Tzeng,~\IEEEmembership{Fellow,~IEEE,}
            and Rajkumar Buyya,~\IEEEmembership{Fellow,~IEEE} 
\IEEEcompsocitemizethanks{\IEEEcompsocthanksitem Xiangbo Li is with Brightcove Inc. E-mail: xli@brightcove.com
\IEEEcompsocthanksitem Magdy Bayoumi, and Nian-Feng Tzeng are with the Center for Advanced Computer Studies, University of Louisiana at Lafayette, LA 70503, USA.\protect\\
 E-mail: \{mab, tzeng\}@cacs.louisiana.edu
\IEEEcompsocthanksitem Mohsen Amini Salehi is with the HPCC lab., School of Computing and Informatics, University of Louisiana at Lafayette, LA 70503, USA.\protect\\
 E-mail: amini@louisiana.edu
 \IEEEcompsocthanksitem Rajkumar Buyya is with the Department of Computing and Information Systems, The University of Melbourne, Melbourne, VIC 3010, Australia.\protect\\
 E-mail: rbuyya@unimelb.edu.au
 }
 
\thanks{}
}


\IEEEcompsoctitleabstractindextext{

\begin{abstract}
Video streams, either in the form of Video On-Demand (VOD) or live streaming, usually have to be converted (\ie transcoded) to match the characteristics of viewers' devices (\eg in terms of spatial resolution or supported formats). Transcoding is a computationally expensive and time-consuming operation. Therefore, streaming service providers have to store numerous transcoded versions of a given video to serve various display devices. With the sharp increase in video streaming, however, this approach is becoming cost-prohibitive. Given the fact that viewers' access pattern to video streams follows a long tail distribution, for the video streams with low access rate, we propose to transcode them in an on-demand (\ie lazy) manner using cloud computing services. The challenge in utilizing cloud services for on-demand video transcoding, however, is to maintain a robust QoS for viewers and cost-efficiency for streaming service providers. To address this challenge, in this paper, we present the Cloud-based Video Streaming Services (\name) architecture. It includes a QoS-aware scheduling component that maps transcoding tasks to the Virtual Machines (VMs) by considering the affinity of the transcoding tasks with the allocated heterogeneous VMs. To maintain robustness in the presence of varying streaming requests, the architecture includes a cost-efficient VM Provisioner component. The component provides a self-configurable cluster of heterogeneous VMs. The cluster is reconfigured dynamically to maintain the maximum affinity with the arriving workload. Simulation results obtained under diverse workload conditions demonstrate that \name architecture can maintain a robust QoS for viewers while reducing the incurred cost of the streaming service provider by up to 85\%.
\end{abstract}

}

\maketitle

\IEEEdisplaynotcompsoctitleabstractindextext
\IEEEpeerreviewmaketitle

\section{Introduction}\label{sec:intro}

The way people watch videos has dramatically changed over the past years. From traditional TV systems, to video streaming on desktops, laptops, and smart phones through the Internet. Consumer adoption of video streaming services is rocketing. Based on the Global Internet Phenomena Report~\cite{intro_1}, video streaming currently constitutes approximately 64\% of all U.S. Internet traffic. It is estimated that streaming traffic will increase up to 80\% of the whole Internet traffic by 2019\cite{intro_2}.

Video contents, either in the form of Video On Demand (VOD) (\eg YouTube\footnote{https://www.youtube.com} or Netflix\footnote{https://www.netflix.com}) or live-streaming (\eg Livestream\footnote{https://livestreams.com}), need to be converted based on the device characteristics of viewers. That is, the original video has to be converted to a supported resolution, frame rate, video codec, and network bandwidth to match the viewers' devices~\cite{intro_6}.  The conversion is termed \textit{video transcoding}~\cite{intro_7}, which is a computationally heavy and time-consuming process~\cite{intro_6}. 
One approach currently used by streaming providers for transcoding is termed \emph{pre-transcoding}, in which several transcoded versions of a given video are stored to serve different types of devices. However, this approach requires massive storage and processing resources. In addition, recent studies (\eg~\cite{intro_9}) reveal that the access pattern to video streams follows a long tail distribution. That is, there is a small percentage of videos that are accessed frequently while the majority of them are accessed very infrequently. Therefore, with the explosive demand for video streaming and the large diversity of viewing devices, the \emph{pre-transcoding} approach is inefficient.

In this research, we propose to transcode the infrequently accessed video streams in an \emph{on-demand} (\ie lazy) manner using computing services offered by cloud providers. 


The challenge for on-demand video transcoding is how to utilize cloud services to maintain a robust Quality of Service (QoS) for viewers, while incurring the minimum cost to the Streaming Service Provider (SSP).

Video stream viewers have unique QoS demands. In particular, they need to receive video streams without any delay. Such delay may occur either during streaming, due to an incomplete transcoding task by its presentation time, or at the beginning of a video stream. In this paper, we refer to the former delay as \emph{missing presentation deadline} and the latter as the \emph{startup delay} for a video stream. Previous studies (\eg \cite{intro_9}) confirm that viewers mostly do not watch video streams to the end. However, they rank the quality of a stream provider based on the video stream's startup delay. Another reason for the importance of the startup delay is the fact that once the beginning part of a stream is processed and buffered, the provider has more time to process the rest of the video stream.
Therefore, to maximize viewers' satisfaction, we define viewers' QoS demand as: \emph{minimizing the startup delay and the presentation deadline violations}. 

To minimize the network delay, transcoded streams are commonly delivered to viewers through Content Delivery Networks (CDNs)~\cite{intro_12}. It is worth noting that, this research is not about the CDN technology. Instead, it concentrates on the computational and cost aspects of on-demand video transcoding using cloud services.

The goal of SSPs is to spend the minimum for renting cloud services, while maintaining a robust QoS for viewers. To satisfy this goal, in our earlier work~\cite{pre_3}, we investigated using homogeneous cloud Virtual Machines (VMs). One extension, we propose in this work, is to consider the fact that cloud providers offer heterogeneous types of VMs. For instance, Amazon EC2 provides General Purpose, CPU-Optimized, GPU-Optimized, Memory-Optimized, Storage-Optimized, and Dense-Storage VMs\footnote{https://aws.amazon.com/ec2/instance-types} with costs varying significantly. Moreover, the execution time of different transcoding operations varies on different VM types. That is, different transcoding operations have different affinities with different VM types.
The challenge is how to construct a heterogeneous cluster of VMs to minimize the incurred cost of SSPs while the QoS demands of viewers are respected? More importantly, the heterogeneous VM cluster should be self-configurable. That is, based on the arriving transcoding tasks, the \emph{number} and the \emph{type} of VMs within the cluster should be dynamically altered to maximize the affinity with VMs and reduce the incurred cost. 

Based on aforementioned definitions, the specific research questions we address in this article are:
\begin{itemize}
 \item How can SSPs satisfy the QoS demands of viewers by minimizing both the video streaming startup delay and presentation deadline violations?
 \item How can SSPs minimize their incurred costs through utilizing a self-configurable heterogeneous VM cluster while maintaining a robust QoS for the viewers?
\end{itemize} 

Previous works (\eg~\cite{rw_3,rw_12}) either did not consider on-demand transcoding of video streams or disregarded the specific QoS demands. Therefore, to answer these research questions, we propose the \textbf{C}loud-based \textbf{V}ideo \textbf{S}treaming \textbf{S}ervice (\name) architecture that enables on-demand video transcoding using cloud services. The architecture includes a scheduling component that maps transcoding tasks to cloud VMs with the goal of satisfying viewers' QoS demands. It also includes a VM Provisioner component that minimizes the incurred cost of the SSP through constructing a self-configurable heterogeneous VM cluster, while maintaining robust QoS for viewers.

In summary, the key \textbf{contributions} of this paper are as follows:
\begin{itemize}
 \item Proposing the \name architecture that enables on-demand transcoding of video streams.
 \item Developing a QoS-aware scheduling component within the \name architecture to map the transcoding tasks to a heterogeneous VM cluster with respect to the viewers' QoS demands.
 \item Developing a VM Provisioner component within the \name architecture that forms a self-configurable heterogeneous VM cluster to minimize the incurred cost to the SSPs while maintaining a robust QoS for viewers.
 \item Analyzing the behavior of the \name architecture from the QoS, robustness, and cost perspectives under various workload intensities. 
\end{itemize}

The rest of the paper is organized as follows. Section~\ref{sec:bg} provides a background on video streaming and transcoding. In Section~\ref{sec:sm}, we present the \name architecture. The scheduling and the VM provisioning policies will be discussed in Sections~\ref{sec:ts} and~\ref{sec:rap}, respectively. In Section~\ref{sec:pe}, we perform performance evaluations. Section~\ref{sec:rw} discusses related works in the literature, and finally Section~\ref{sec:conclusion} concludes the paper and provides avenues of future work. 

\section{Background}\label{sec:bg}
\subsection{Definition of Robustness}
\emph{Robustness} is defined as the degree to which a system can function correctly in the presence of uncertain parameters in the system~\cite{rw_17}. 

In a system for on-demand transcoding, the arrival pattern of the streaming requests is uncertain, which can significantly harm QoS and viewer satisfaction~\cite{intro_13}. Ideally, the system has to be robust against uncertainty in the arrival pattern of the streaming requests. That is, the system has to satisfy a certain level of QoS, even in the presence of uncertain arrival of streaming requests.

\subsection{Video Stream Structure}\label{vbs}
\begin{figure}[htb] 
    \centering
    \includegraphics[width=0.8\columnwidth]{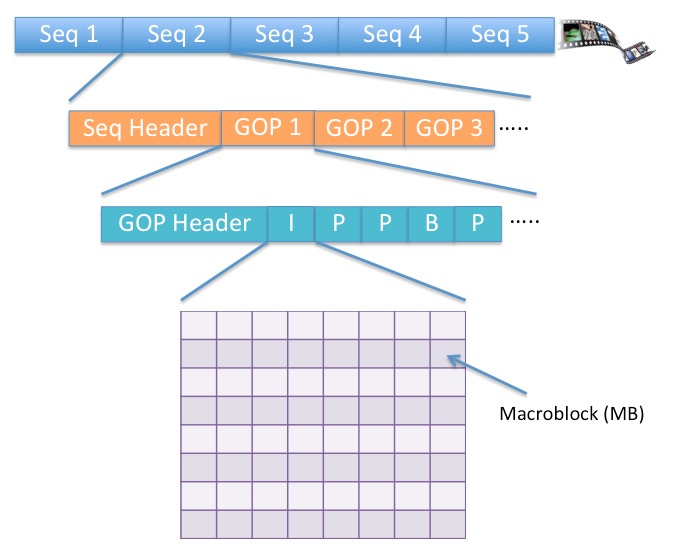}
    \caption{The structure of a video stream. It consists of several sequences. Each sequence includes multiple GOPs. Each frame of a GOP contains several MacroBlocks.}
    \label{fig:video_segment}
\end{figure}

A Video stream, as shown in Figure~\ref{fig:video_segment}, consists of several sequences. Each sequence is further divided into multiple \textit{Group Of Pictures} (GOPs) with sequence header information at the beginning.
Each GOP essentially comprises a sequence of frames beginning with an \texttt{I} (intra) frame, followed by a number of \texttt{P} (predicted) frames or \texttt{B} (bi-directional predicted) frames. 
Each frame of a GOP contains several \textit{slices} that consist of a number of \textit{macroblocks} (MB) which is used for video encoding and decoding. 
In practice, video streams are commonly split into \emph{GOP tasks} (simply termed GOPs in the paper) for processing that can be transcoded independently~\cite{bg_2}. 

\subsection{Video Transcoding}\label{sec:vt}
A video initially is captured with a particular format, spatial resolution, frame rate, and bit rate. Then, the video is uploaded to a streaming server where it is adjusted based on the viewer's device resolution, frame rate, and video codec. These conversions are generally termed \textit{video transcoding}~\cite{intro_6, intro_7} operations and are explained as follows:

\noindent\textbf{Bit Rate Adjustment.}
To produce a high quality video contents, the video is encoded with high bit rate. However, a higher bit rate also requires larger network bandwidth for video stream transmission. SSPs usually need to transcode the video stream to adjust the bit rate based on available viewer bandwidth~\cite{bg_3}.

\noindent\textbf{Spatial Resolution Reduction.}
Spatial resolution indicates the encoded dimensional size of a video. However, the dimensional size does not necessarily match the screen size of the viewer's device. To avoid losing contents, macroblocks of an original video have to be removed or combined (\ie downscaled) to produce a lower spatial resolution video~\cite{bg_7}. 
 

\noindent\textbf{Temporal Resolution Reduction.}
Temporal resolution reduction happens when the viewer's device only supports a lower frame rate, and hence, some frames have to be dropped. Due to dependency between frames, dropping frames can invalidate motion vectors (MV) for the incoming frames. Temporal resolution reduction can be achieved using methods explained in~\cite{bg_12}.

\noindent\textbf{Compression Standard (Codec) Conversion.}
Video compression standards vary from MPEG2 to H.264, and to the most recent one, HEVC. MPEG2 is widely used for DVD and video broadcasting, while HD or Blu-ray videos are mostly encoded with H.264. HEVC is the latest and most efficient compression standard. Viewer devices usually support a specific codec. Thus, video streams need to be transcoded from the original codec to the one supported by the viewer's device~\cite{bg_10}.

\subsection{Video Transcoding Using Heterogeneous VMs }\label{subsec:pricing}
Cloud providers usually offer numerous VM types. For instance, Amazon EC2 currently provides more than 40 VM types. These VM types are heterogeneous both in terms of their underlying hardware architectures and prices. 
In Amazon EC2, VMs are categorized in 6 groups based on their architectural configurations. In particular, these groups are: General-Purpose, CPU-Optimized, Memory-Optimized, GPU-Optimized, Storage-Optimized, and Dense-Storage. 

\begin{figure}[htb] 
    \centering
    \includegraphics[width=1.0\columnwidth]{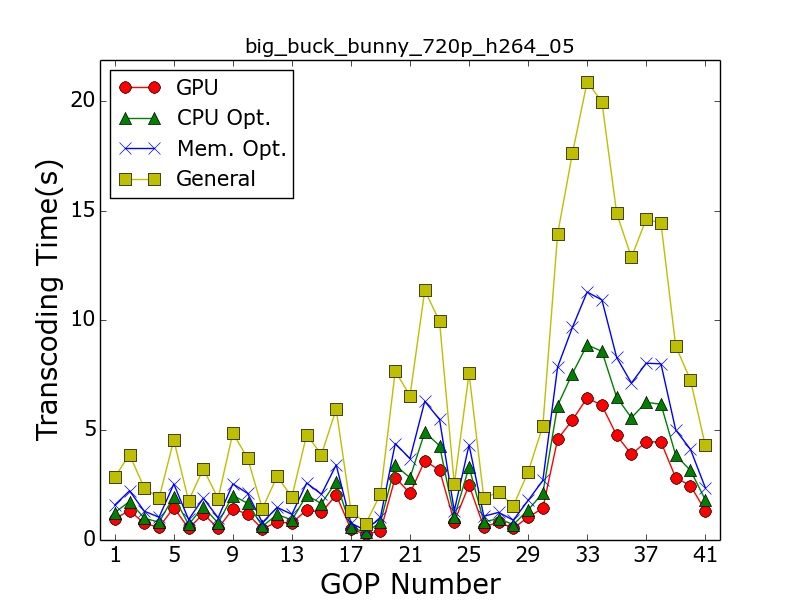}
    \caption{Transcoding time (in seconds) of GOPs using different VM types. The horizontal axis shows the sequential order of GOP numbers in a video stream.}
    \label{fig:het_codec}
\end{figure}

Our initial evaluations on transcoding the codec of a set of benchmark videos\footnote{the workload trace of the benchmark videos are available from}: \url{https://goo.gl/B6T5aj} (explained in Section~\ref{subsec:es}) demonstrated that transcoding GOPs have different execution times on various VM types. In particular, we executed GOPs on four VM types, and their performance results are shown in Figure~\ref{fig:het_codec}\footnote{Figure~\ref{fig:het_codec} shows the result for one of the benchmark videos. We used big buck bunny 720p in the benchmark for this experiment. However, results for other experiments confirm the same observations.}. We did not consider any of the Storage Optimized and Dense Storage VM types in our evaluations as we observed that IO and storage are not influential factors for transcoding tasks. 
Due to huge diversity, we selected one VM instance that represents the characteristics within each category. More specifically, for GPU instance, CPU-Optimized, Memory-Optimized, and General-Purpose types we chose \texttt{g2.2xlarge}, \texttt{c4.xlarge}, \texttt{r3.xlarge}, and \texttt{m4.large}, respectively. The cost of the chosen instance types are illustrated in Table~\ref{tbl:cost-table}.

The vertical axis of Figure~\ref{fig:het_codec} shows the transcoding time (\ie execution time) for different GOPs of a given video stream. According to the figure, in general, GPU instances provide a lower execution time than other VM instance types. However, for some of the GOPs, the performance difference of GPU with other VM instances is negligible, while its cost is remarkably higher (see Table~\ref{tbl:cost-table}). 
The experiment indicates that an SSP can utilize heterogeneous VM types to minimize its incurred cost while satisfying viewers' QoS demands. 

\begin{table}[htb]
  \caption{\label{tbl:cost-table}Cost of different VM types in Amazon EC2}
  \centering{
    \resizebox{\columnwidth}{!}{
      \begin{tabular}{ c | c | c | c | c }
       VM Type & \shortstack{GPU\\ (\texttt{g2.xlarge})} & \shortstack{CPU Opt.\\ (\texttt{c4.xlarge})} & \shortstack{Mem. Opt.\\ (\texttt{r3.xlarge})} & \shortstack{General \\ (\texttt{m4.large})} \\ \hline
       Hourly Cost (\$) & 0.65 & 0.20 & 0.33 & 0.15 \\ 
      \end{tabular}
    }
  }
\end{table}

\section{\name: Cloud-Based Video Streaming Service Architecture}\label{sec:sm}
\subsection{Overview}
\begin{figure}[htb] 
    \centering
    \includegraphics[width=0.85\columnwidth]{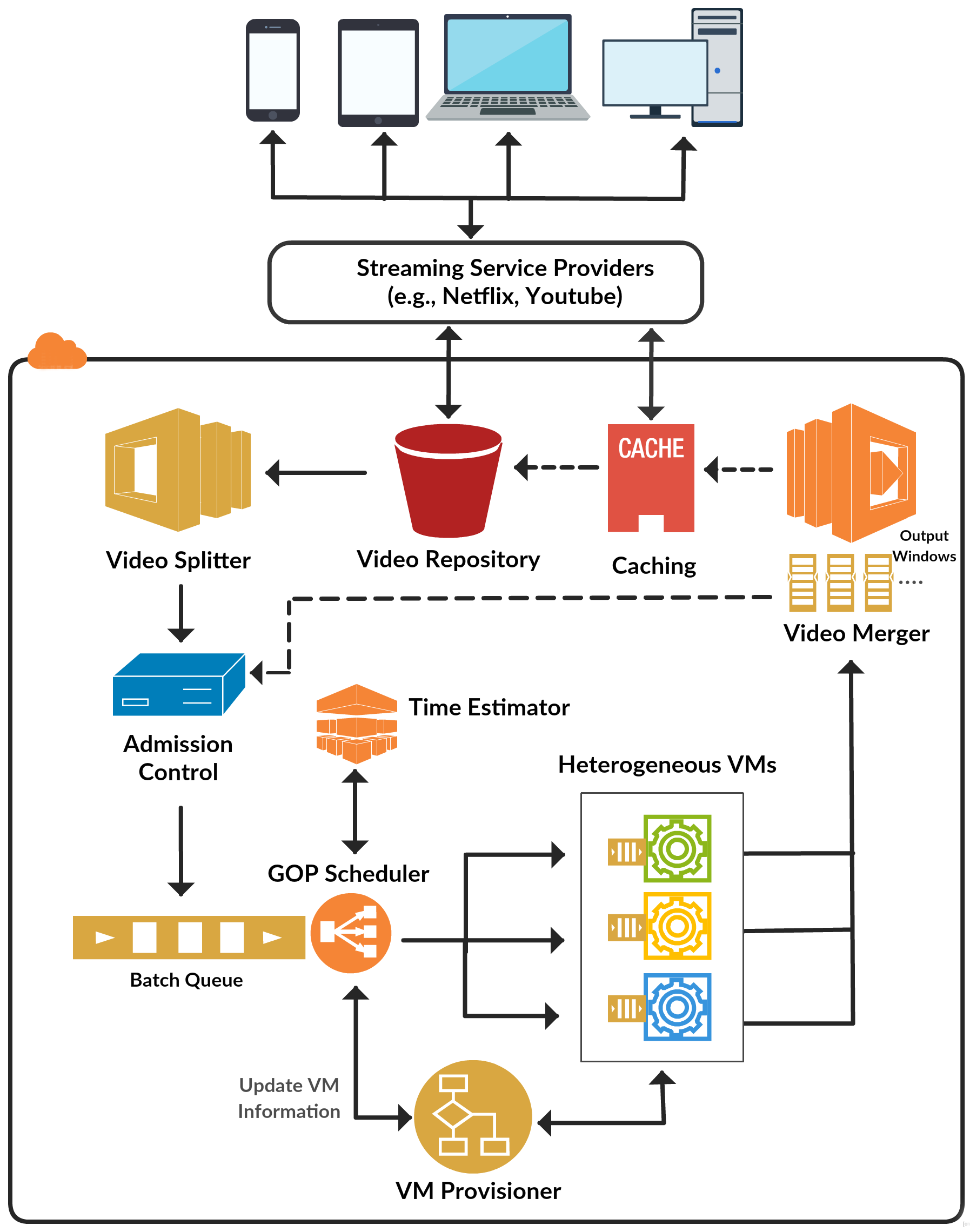}
    \caption{An overview of the Cloud-based Video Streaming Service (\name) architecture.}
    \label{sa}
\end{figure}

The \name architecture aims to deal with a received request for streaming a video format that is not available in the repository (\ie it is not pre-transcoded). An overview of the architecture is presented in Figure~\ref{sa}. It shows the sequence of actions taken place to transcode a video stream in an on-demand manner. The dashed lines in this figure will be investigated in our future studies.

\name architecture includes eight main components, namely \textit{Video Splitter, Admission Control, Time Estimator, Task (\ie GOP) Scheduler, Heterogeneous Transcoding VMs, VM Provisioner, Video Merger,} and \textit{Caching}. These components are explained in the next few subsections. 

\subsection{Video Splitter}\label{sec:vs}
The Video Splitter splits the video stream into several GOPs that can be transcoded independently. Each generated GOP is identified uniquely in form of $G_{ij}$, where $i$ is the video stream id and $j$ is the GOP number within the video stream. 

Each GOP is treated as a task with an individual deadline. The deadline of a GOP is the presentation time of the first frame in that GOP. In the case of VOD, if a GOP misses its deadline, it still has to complete its transcoding. We have made the source code for Video Splitter publicly\footnote{The source code for GOP task generation is available here:\url{https://github.com/lxb200709/videotranscoding_gop}} available.

\subsection{Admission Control}\label{sec:adc}
The Admission Control component includes policies that regulate GOP dispatching to the scheduling queue. 
In fact, the Video Splitter generates GOPs for all requested video streams. Then, the admission control policies determine the priority (\ie urgency) of the GOPs and dispatches them accordingly to the scheduling queue. 
The admission control policies act based on the inputs it receives from Video Splitter and Video Merger. 

The way Admission Control prioritizes a GOP is based-on the GOP sequence number in a video stream. Details of how to prioritize GOP tasks is explained in Section~\ref{subsec:utility}

\subsection{Transcoding Virtual Machines (VMs)}\label{sec:vmts}
VMs are allocated from the cloud provider to transcode GOP tasks. 
As discussed in Section~\ref{subsec:pricing}, cloud providers offer VMs with diverse architectural configurations. Although GOPs can be processed on all VM types, their execution times vary. In fact, the execution time of a GOP on a particular VM type can depend on factors such as the size of data it processes or the type of transcoding operations it performs. 


Each VM is assigned a local queue where the required data for GOPs are preloaded before execution. The scheduler maps GOPs to VMs until the local queue gets full.  
\subsection{Execution Time Estimator}\label{sec:te} 
The role of the Time Estimator component is to estimate the execution time of GOP tasks. Such estimation of execution times helps the Scheduler and VM Provisioner components to function efficiently.  

In VOD streaming, a video usually has been streamed multiple times. 
Therefore, the transcoding execution time for each $G_{ij}$ 
can be estimated from the historic execution information of $G_{ij}$~\cite{sm_1}. 

As we consider the case of heterogeneous transcoding VMs, each GOP has a different execution time on each VM type. Therefore, the Time Estimator stores the execution time estimations within Estimated Time to Completion (ETC) matrices~\cite{rw_17}. An entry of the ETC matrix expresses the execution time of a given GOP $G_{ij}$ on a given VM type $m$.

We note that, even in transcoding the same GOP $G_{ij}$ on the same type of VM, there is some randomness (\ie uncertainty) in the transcoding execution time. That is, the same VM type does not necessarily provide identical performance for executing the same GOP at different times~\cite{sch_2}. This variance is attributed to the fact that the same VM type can be potentially allocated on different physical machines on the cloud. It can also be attributed to other neighboring VMs that coexist with the VM on the same physical host in the cloud datacenter. For instance, if the neighboring VMs have a lot of memory access, then, there will be a contention to access the memory and the performance of the VM will be different from the situation that there is no such a neighboring VM. Therefore, to capture randomness that exists in the GOP execution time, the mean execution time and its standard deviation of the historic execution time for $G_{ij}$ is stored in the corresponding entry of the ETC matrix.


\subsection{Transcoding (GOP) Task Scheduler}\label{sec:tts}
The GOP task scheduler (briefly called transcoding scheduler) is responsible for mapping GOPs to a set of heterogeneous VMs. Considering the heterogeneity in performance and cost of different VM types, the scheduler's goal is to map GOP tasks to VMs with the minimum incurred cost while satisfying the QoS demands of the viewers. 

GOPs of different video streams are interleaved within the scheduling queue. In addition, the scheduler has no prior knowledge about the arrival pattern of the GOPs to the system. Details of the scheduling method are presented in Section~\ref{sec:ts}.

\subsection{VM Provisioner}\label{sec:vcm} 
The VM Provisioner component monitors the operation of transcoding VMs in the \name architecture and dynamically reconfigures the VM cluster with two goals: (A) minimizing the incurred cost to the stream provider; (B) maintaining a robust QoS for viewers. For that purpose, the VM Provisioner includes provisioning policies that are in charge of \emph{allocating} and \emph{deallocating} VM(s) from the cloud based on the streaming demand type and rate. 


VM provisioning policies generally have to determine \emph{when} and \emph{how many} VMs need to be provisioned (known as elasticity~\cite{pre_3}). For a heterogeneous VM cluster, the policy also has to determine \emph{which type} of VM needs to be provisioned. 

The VM provisioning policies are executed periodically and also in an event-based fashion to verify whether or not the allocated VMs are sufficient to meet the QoS demands. Once the provisioning policy updates the set of allocated VMs, it informs the scheduler about the latest configuration of the VM cluster. 
Details of the VM provisioning policies are discussed in Section~\ref{sec:rap}. 
\subsection{Video Merger}\label{sec:vm}
GOPs are transcoded on different VMs independently. Thus, latter GOPs in a video stream may be completed before the earlier ones in a stream. The role of Video Merger is to rebuild the sequence of GOPs in the right order. To build the transcoded stream, Video Merger maintains an output window for each video stream.  

Video Merger is in contact with the Admission Control component. In the event that a GOP is delayed (\eg due to failure) the Video Merger asks the Admission Control for resubmission of the GOP. Upon receiving a resubmission request, Admission Control fetches the requested GOP from Splitter and resubmits it to the Scheduler with a high priority. 

Video Merger requests for resubmission of a GOP after a certain time elapsed and it does not need to search for the missed GOP to see if it has failed or not.


\subsection{Caching}\label{sec:cp}

To avoid redundant transcoding of the trending videos, the \name architecture provides a caching policy to decide whether a transcoded video should be stored or not. If the video is barely requested by viewers, there is no need to store (\ie cache) the transcoded version. Such videos are transcoded in an on-demand manner upon viewers' request. We will explore more details of the caching policy in a future research.

Considering the proposed architecture, in the next two sections, we elaborate on the methods developed for the Transcoding Task Scheduler and VM Provisioner components.
\section{QoS-Aware Transcoding (GOP) Task Scheduler}\label{sec:ts}
\subsection{Overview}

\begin{figure}[htb] 
    \centering
    \includegraphics[width=0.8\columnwidth]{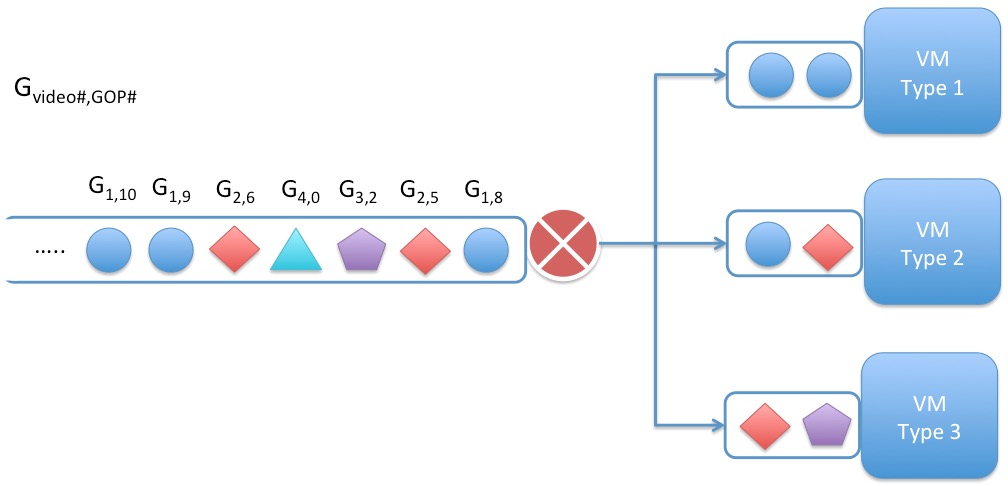}
    \caption{QoS-aware transcoding scheduler that functions based on the utility value of the GOPs.}
    \label{fig:sche}
\end{figure}

Details of the GOP task scheduler are shown in Figure~\ref{fig:sche}. 
According to the scheduler, GOPs of the requested video streams are batched in a queue upon arrival to be mapped to VMs by the scheduling method. To avoid any execution delay, the required data for GOPs are fetched in the local queue of the VMs, before the GOP transcoding started. Previous studies~\cite{rw_17} show that the local queue size should be short. Accordingly, we consider the local queue size to be 2 in all VMs. We assume that the GOP tasks in the local queue are scheduled in the \textit{first come first serve} (FCFS) fashion. Once a free slot appears in a VM local queue, the scheduling method is notified to map a GOP task from those in the batch queue to the free slot. We assume that GOP scheduling is non-preemptive and non-multi-tasking. 

Recall that the scheduler goal is to satisfy the QoS demands of viewers by minimizing the average deadline miss rate and the average startup delay of the video streams. 
The scheduling method maps the GOP tasks to a heterogeneous cluster of VMs where GOPs have different execution times on different VM types. In such a system, optimal mapping of GOP tasks to heterogeneous VMs is an NP-complete problem~\cite{sch_8}. Thus, development of mapping heuristics to find near-optimal solutions forms a large body of research~\cite{rw_9, rw_17}. 

In the rest of this section, we explain the details of how the scheduling component within the \name architecture satisfies the QoS demands. Also, for further clarity, all the symbols used in this paper are listed in Table~\ref{tab:TableOfNotationForMyResearch}, in Appendix~\ref{apdx:tbl} Section.  

\subsection{Utility-based GOP Task Prioritization}\label{subsec:utility}
One approach to minimize the average startup delay of video streams is to consider a separate dedicated queue for the startup GOPs of the streams~\cite{pre_3}. Such a queue can only prioritize a constant number of GOPs at the beginning of the streams, with the rest of the GOPs treated as normal priority.  
In practice, however, the priority of GOPs should be decreased gradually as the video stream moves forward.
 
To implement the gradual prioritization of GOPs in a video stream, we define a \emph{utility function} that operates on a video stream and assigns \emph{utility values} to each GOP. Equation (\ref{eq:utility}) shows the utility function the admission control policy uses for assigning utility values. 
In Equation (\ref{eq:utility}), $c$ is a constant and $i$ is the order number of GOP in the video stream. The value of $c$ determines the slope of the utility function curve. That means, using this parameter we can adjust the importance of the startup GOPs in a video stream. Higher values for $c$ create a sharp slope in the curve that implies prioritizing few GOPs in the beginning of the video stream with a high utility value and low utility values for the rest of GOPs in the video stream. Our initial experiments showed that $c = 0.1$ provides a reasonable slope in Equation (\ref{eq:utility}).That is, it assigns a high utility value to the GOPs in the beginning of the stream and then the utility value gradually decreases for GOPs positioned later in the stream.

\begin{equation}\label{eq:utility}
U_i = (\frac{1}{e})^{c \cdot i} 
\end{equation}

\begin{figure}[htb] 
    \centering
    \includegraphics[width=0.7\columnwidth]{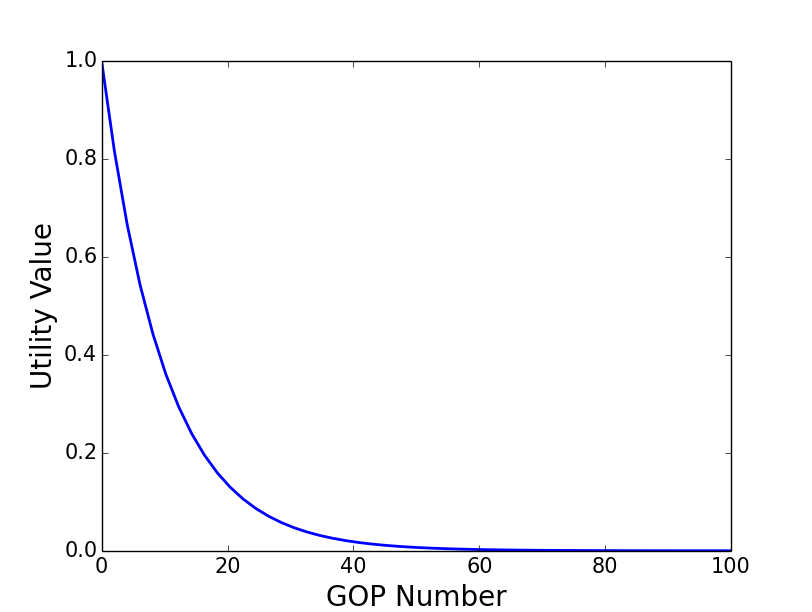}
    \caption{Utility values of different GOP tasks to indicate their processing priority within a video stream. }
    \label{fig:priority}
\end{figure}

The utility values assigned to a given video stream are depicted in Figure~\ref{fig:priority}. In this figure, the horizontal axis is the GOP number and the vertical axis is the utility value. As we can see, the utility function assigns higher utility values (\ie higher priority) to earlier GOPs in the stream. The utility value drops for the latter GOPs in the stream. 

We would like to note that, although we used Equation (\ref{eq:utility}) to assign utility values to GOP tasks, our proposed method is general and its operation is not dependent on this particular utility function. In fact, our proposed methods can operate under any utility function as long as it assures that the first part of the video is prioritized more than the rest of it.

\subsection{Estimating Task Completion Time on Heterogeneous VMs}
For each GOP $j$ from video stream $i$, denoted $G_{ij}$, the arrival time and the deadline (denoted $\delta_{ij}$) are available. It is worth noting that the GOP deadline is relative to the beginning of the video stream. Therefore, to obtain the absolute deadline for $G_{ij}$ (denoted  $\Delta_{ij}$) the relative deadline must be added to the presentation start time of the video stream (denoted $\psi_{i}$). That is, $\Delta_{ij} = \delta_{ij} + \psi_i$.

Recall that the estimated execution time for $G_{ij}$ on VM type $m$ is available through the ETC matrix (see Subsection~\ref{sec:te}).
To capture randomness in the estimated execution time of GOPs, let $\tau^m_{ij}$ be the worst-case transcoding time estimation. That is, in the scheduling, we consider $\tau^m_{ij}$ as the sum of mean historic execution times of $G_{ij}$ and its standard deviation on $VM_m$.  

Our scheduling method also needs to estimate the tasks' \textit{completion times} to be able to efficiently map them to VMs. To estimate the completion time of an arriving GOP task $G_n$ on $VM_m$, we add up the estimated remaining execution time of the currently executing GOP in $VM_m$ with the estimated execution time of all tasks ahead of $G_n$ in the local queue of $VM_m$. Finally, we add the estimated execution time of $G_n$ (\ie $\tau^m_n$). Recall that each GOP task has a different execution time on different VM types that can be obtained from the ETC matrix (see Section~\ref{sec:te}).
Let $t_r$ denote the remaining estimated execution time of the currently executing task on $VM_m$, and let $t_c$ be the current time. Then, we can estimate the \emph{task completion time} of $G_n$ on $VM_m$ (denoted $\varphi^m_n$) as follows:

\begin{equation}\label{eq:cmpl}
\varphi^m_n = t_c + t_r + \sum_{p=1}^{N} \tau^m_p + \tau^m_n
\end{equation}

where $\tau^m_p$ denotes the worst case estimated execution time of any task waiting ahead of $G_n$ in local queue of $VM_m$ and $N$ is the number of waiting tasks in local queue of $VM_m$. 

\subsection{Mapping Heuristics}\label{subsec:mapheur}
Mapping heuristics are responsible to map tasks from the batch queue to machine queues (see Figure~\ref{fig:sche}).Regardless of their implementation details, mapping heuristics for heterogeneous computing systems have a general mechanism that operates in two main phases~\cite{sch_9}. In Phase 1, for all tasks in the batch queue, the machine (\ie VM) that provides the minimum expected completion time is determined. The output of this phase can be considered as pairs of tasks with the machines that provide the minimum expected completion time for them. Then, in Phase 2, from the set of task-machine pairs identified in Phase 1, the mapping heuristic selects the pair that maximizes its performance objective. This process is repeated until either all tasks in the batch queue are assigned or there is no free slot left in machine queues. 

Based on the explained mechanism, MinCompletion-MinCompletion (MM)~\cite{sch_3,sch_4,sch_5,sch_6}, MinCompletion-SoonestDeadline (MSD)~\cite{rw_17,sch_7}, and MinCompletion-MaxUrgency (MMU)~\cite{rw_17,sch_7} mapping heuristics are defined as follows:
 
\noindent\textbf{MinCompletion-MinCompletion (MM):} 
In Phase 1, the heuristic finds the machine (\ie VM) that provides the minimum expected completion time for the GOP task. In Phase 2, the heuristic selects the pair that has the minimum completion time from all the task-machine pairs generated in the Phase 1. Once the selected task is mapped to the selected machine, it is removed from the batch queue.

\noindent\textbf{MinCompletion-SoonestDeadline (MSD)} 
In Phase 1, for each task in the batch queue, the heuristic finds the VM that provides the minimum expected completion time. In Phase 2, from the list of task-machine pairs found in the Phase 1, MSD assigns the task that has the soonest deadline. 
 
\noindent\textbf{MinCompletion-MaxUrgency (MMU):} 
In Phase 1 of MMU, for each task in the batch queue, the heuristic finds the VM that provides the minimum expected completion time. In Phase 2, from the list of task-machine pairs found in the Phase 1, MMU assigns the task whose task urgency is the greatest (\ie has the shortest slack). 

Although these mapping heuristics are extensively employed in heterogeneous computing systems, none of them consider the task precedence based on the utility value as discussed in Section~\ref{subsec:utility}. 

\subsubsection{Utility-Based Mapping Heuristics}

\begin{figure}[htb] 
    \centering
    \includegraphics[width=0.8\columnwidth]{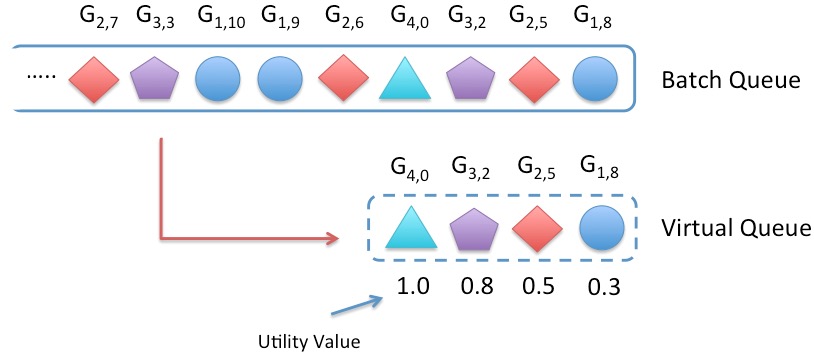}
    \caption{Virtual Queue to hold GOPs with the highest utility values from different video streams. GOPs in Virtual Queue are ready for mapping to VMs.}
    \label{fig:sche_virtual}
\end{figure}

Recall that each GOP is assigned a utility value that shows its precedence. Therefore, in the \emph{first phase} of our proposed scheduling method, as shown in Figure~\ref{fig:sche_virtual}, the GOPs with the highest utility values are selected and put into a virtual queue. The rest of the scheduling method is applied on the virtual queue rather than the whole batch queue. Given the large number of GOPs in the batch queue, making use of the virtual queue reduces the scheduling overhead. 

In the \emph{second phase}, similar to the heuristics introduced in Subsection~\ref{subsec:mapheur}, task-VM pairs are formed based on the VM that provides the minimum expected completion time for each GOP in the priority queue. Then, in the \emph{third phase}, the mapping decision is made by combining a performance objective (\eg  SoonestDeadline) and the utility values of the GOP tasks. For combining, we prioritize the GOP with the highest utility value from the pairings of a VM, if and only if it does not violate the deadline of the task selected based on the performance objective. 

To clarify further, we explain the third phase using an example. Let GOP tasks $G_a$ and $G_b$ denote pairs for $VM_m$. Also, let SoonestDeadline be the performance objective. Assume that $G_a$ has a sooner deadline, whereas $G_b$ has a higher utility value. In this case, $G_b$ can be assigned to $VM_m$, if and only if it does not violate the deadline of $G_a$. 
To assure that assigning $G_b$ does not cause a violation of the deadline of $G_a$, we assume that $G_b$ has already been assigned to $VM_m$ and run the mapping heuristic again to see if $G_a$ can still meet its deadline or not.

Based on the way the third phase of our proposed mapping heuristic functions, we can have 3 variations, namely Utility-based MinCompletion-MinCompletion (MMUT), Utility-based MinCompletion-SoonestDeadline (MSDUT), and Utility-based MinCompletion-MaximumUrgency (MMUUT). 

\section{Self-Configurable Heterogeneous VM Provisioner}\label{sec:rap}
\subsection{Overview}
The goal of the VM Provisioner component is to maintain a robust QoS while minimizing the incurred cost to the stream provider. To that end, the component includes VM provisioning policies that make decisions for \emph{allocating} and \emph{deallocating} VMs from cloud. 

To achieve the QoS robustness, the SSP needs to define the acceptable QoS boundaries. Therefore, the SSP provides an upper bound threshold for the deadline miss rate of GOPs that can be tolerated, denoted $\beta$. Similarly, it provides a lower bound threshold for the deadline miss rate, denoted $\alpha$, that enables the provisioning policies to reduce the incurred cost of the stream provider through deallocating VM(s). 

The strategy of the VM provisioning to maintain QoS robustness is to manage the VM allocation/deallocation so that the deadline miss rate at any given time $t$, denoted $\gamma_{t}$, remains between $\alpha$ and $\beta$. That is, at any given time $t$, we should have $\alpha \leq \gamma_t \leq \beta$. 

The VM Provisioner component follows the \emph{scale up early and scale down slowly} principle. That is, VM(s) are allocated from the cloud as soon as a provisioning decision is made. However, as the stream provider has already paid for the current charging cycle of the allocated VMs, the deallocation decisions are not practiced until the end of the current charging cycle.
  
In general, any cloud-based VM provisioning policy needs to deal with two main questions:
\begin{enumerate}
\item  \emph{When} to provision VMs?
\item \emph{How many} VMs to provision?
\end{enumerate} 
The self-configurable VM provisioning, however, introduces a third question to the VM provisioning policies:
\begin{enumerate}
  \setcounter{enumi}{2}
  \item \emph{What type} of VM(s) to provision?
\end{enumerate} 

In the next subsections, we first provide a method to determine the suitability of VM types for GOP tasks,  then we introduce two provisioning policies, namely periodic and remedial, that work together to answer the three aforementioned questions.

\subsection{Identifying Suitability of VM Types for GOP Tasks}\label{subsec:goptype}

Recall that each GOP task has different execution times on different VM types (see Section~\ref{sec:bg}). In general, GPU provides a shorter execution time compared with other VM types. However, for some GOPs, the execution time on GPU is close to other VM types while its cost is significantly higher (see Table~\ref{tbl:cost-table}). Therefore, we need a measure to determine the suitability of a VM type for a GOP based on the two factors. 

For a given GOP task, we define \emph{suitability}, denoted $S_i$, as a measure to quantify the appropriateness of a VM type $i$ for executing the GOP task both in terms of performance and cost. We calculate the suitability measure for a task based on Equation (\ref{eq:weight}). The measure establishes a trade-off between the performance ($T_i$) and the cost ($C_i$) for a given GOP on VM type $i$. 
\begin{equation}\label{eq:weight}
      S_i =  k \cdot T_i + (1-k) \cdot C_i
\end{equation}
The value of $k$, in Equation (~\ref{eq:weight}), is determined by the CVS2 user (\ie video stream provider) and represents her preference between performance and cost of VM type $i$. The value of $T_i$ is defined based on Equation (\ref{eq:time}).
\begin{equation}\label{eq:time}
      T_i = \frac{t_{max} - t_i}{t_{max} - t_{min}}
\end{equation}
where $t_i$ is the GOP execution time on VM type $i$ (obtained from the ETC matrix). Also, $t_{max}$ and $t_{min}$ are the maximum and minimum GOP execution times across all VM types, respectively. Nominator of this equation determines the execution time improvement provided by VM type $i$  for the GOP. Denominator of this equation ensures that the value of $T_i$ remains in [0,1] space.

In Equation (\ref{eq:weight}), the value of $C_i$ is determined according to Equation (\ref{eq:cost}).

\begin{equation}\label{eq:cost}
      C_i = \frac{c_{max} - c_{i}}{c_{max} - c_{min}}
\end{equation}
where $c_i$ is the cost of transcoding the same GOP on VM type $i$. Also, $c_{max}$ and $c_{min}$ are the maximum and minimum GOP transcoding costs across all VMs, respectively. The rationale of Equation (~\ref{eq:cost}) is similar to that of Equation (~\ref{eq:time}). Nominator of the equation determines the cost improvement resulted from VM type $i$ to transcode the GOP and denominator ensures the value of $C_i$ remains in [0,1].

Based on Equation (\ref{eq:weight}), for a given GOP task, we define the \emph{GOP type} based on the type of VM that provides the highest suitability value. Later, the VM provisioning policies will utilize the concept of GOP type in their provisioning decisions.
 
\subsection{Periodic VM Provisioning Policy}
This VM provisioning policy occurs periodically (we term it \emph{provisioning event}) to make VM allocation or deallocation decisions. 
The policy includes two methods, namely \emph{Allocation} and \emph{Deallocation}.

\subsubsection{Allocation Method}

Algorithm~\ref{alg:allocation} provides a pseudo-code for the VM allocation method. 
The method is triggered \emph{when} the deadline miss rate ($\gamma_{t}$) goes beyond the upper bound threshold $\beta$ (line 2 in the Algorithm). The value of $\beta$ is determined by the video streaming service provider (\ie CVS2 user) and represents how much the provider can tolerate QoS violation in favor of cost-efficiency.

\begin{algorithm}[htb]
  \caption{Pseudo-code for the VM Allocation Method}
  \label{alg:allocation}
  \begin{algorithmic}[1]
    \INPUT\tikzmark{a} 
    \Statex $\beta$: upper bound threshold for deadline miss rate
    \Statex $r$: streaming request arrival rate \tikzmark{b}
    \OUTPUT\tikzmark{c} 
    \Statex $n$: list of number of VMs of each type to be allocated.\tikzmark{d} 
    \Statex
    \State $\gamma_t \leftarrow$ current deadline miss rate     
    \If{$\gamma_{t} \geq \beta$}
       \For{each VM type $i$}   
          \State $\sigma_i \leftarrow$ deadline miss rate for each GOP type $i$
          \State $\phi_i \leftarrow$ ratio of each GOP type $i$ in the batch queue 
          \State Calculate the demand ($\omega_i$) for each VM type $i$  
          \State $\rho_i \leftarrow$ minimum utilization in VMs of type $i$ 
              \If{$\omega_i \geq \omega_{th}$ and $\rho_i \geq \rho_{th}$}
                \State $n_i \leftarrow \lfloor\frac{r \cdot \omega_i} {\beta}\rfloor$
                \State Allocate $n_i$ VM type $i$
              \EndIf
        \EndFor
    \EndIf  
  \end{algorithmic}
\end{algorithm}

To determine \emph{what type} of VM(s) to be allocated, we need to understand the demand for different VM types. Such demand can be understood from the concept of GOP type, introduced in Subsection~\ref{subsec:goptype}. In fact, the number of GOP tasks from different types can guide us to the types of VMs that are required. More specifically, we can identify the type of required VMs based on two factors: (A) the proportion of deadline miss rate for each GOP type, denoted $\sigma_i$, and (B) the proportion of GOPs of each type waiting for execution in the batch queue, denoted $\phi_i$. In fact, factor (A) indicates the current QoS violation status of the system, whereas factor (B) indicates the QoS violation status of the system in the near future.

Based on these factors, we define the \emph{demand} for each VM type $i$, denoted $\omega_i$, according to Equation (\ref{eq:demand}). The constant factor $0\leq k \leq 1$, in this equation, determines the weight assigned to the current deadline miss rate status and to the future status of the system. 

For implementation, we experimentally realized that the value of $k$ should be determined in a way that GOPs waiting in the batch queue (\ie $\phi_i$) are assigned a higher weight, rather than the current QoS violation of each GOP type (\ie $\sigma_i$). The reason is that, the GOP tasks in the batch queue represent the QoS violation the system will encounter in a near future which is more important than the QoS violation the system currently is encountering. Hence, we considered $k=0.3$ (thus, $1-k=0.7$) in Equation (~\ref{eq:demand}). Based on this justification, we believe that in a system with a different workload scenario than those we considered in our evaluations, the value of $k$ should remain the same.

\begin{equation}\label{eq:demand}
      \omega_i = k \cdot \sigma_i + (1-k) \cdot \phi_i
\end{equation}

If the demand for VM type $i$ is greater than the allocation threshold ($\omega_{th}$ in line 8), and also the utilization of corresponding VM type ($\rho_i$) is greater than the utilization threshold ($\rho_{th}$), then the policy decides to allocate from VM type $i$. 

Once we determine the type of VMs that needs to be allocated, the last question to be answered is \emph{how many} VMs of each type to be allocated (lines 8 - 11 in the Algorithm). The number of allocations of each VM type depends on how far is the deadline miss rate of GOP type $i$ is from $\beta$. For that purpose, we use the ratio of $\omega_i / \beta$ to determine the number of VM(s) of type $i$ that has to be allocated (line 9). The number of VM(s) allocated also depends on the arrival rate of GOP tasks to the system. Therefore, the GOP arrival rate, denoted $r$, is also considered in line 9 of Algorithm~\ref{alg:allocation}.

\subsubsection{Deallocation Method}
The VM deallocation method functions are based on the lower bound threshold ($\alpha$). 
That is, it is triggered \textit{when} the deadline miss rate ($\gamma_{t}$) is less than $\alpha$. Once the deallocation method is executed, it terminates at most one VM. The reason is that, if the VM deallocation decision is practiced aggressively, it can cause loss of processing power and results in QoS violation in the system. Therefore, the only question in this part is \textit{which} VM should be deallocated.

In the first glance, it seems that the deallocation method can simply choose the VM with the lowest utilization for deallocation. However, this is not the case when we are dealing with a heterogeneous VM cluster. The utilizations of the VMs are subject to the degree of heterogeneity in the VM cluster. For instance, when the VM cluster is in a mostly homogeneous configuration, the task scheduler has no tendency to a particular VM type. This causes all VMs in the cluster to have a similar and high utilization. Hence, if the deallocation method functions just based on the utilization, it cannot terminate VM(s) in a homogeneous cluster, even if the deadline miss rate is low.

The challenge is how to identify the degree of heterogeneity in a VM cluster. To cope with this challenge, we need to quantify the VM cluster heterogeneity. Then, we can apply the appropriate deallocation method accordingly. 

We define \emph{degree of heterogeneity}, denoted $\eta$, as a quantity that explains the VM diversity (\ie heterogeneity) that exists within the current configuration of the VM cluster. We utilize the Shannon Wiener equitability~\cite{rp_1} function to quantify the degree of heterogeneity within our VM cluster. The function works based on the Shannon Wiener Diversity Index that is represented in Equation (\ref{eq:diversity_1}).

\begin{equation}\label{eq:diversity_1}
      H = -\sum_{i=1}^N p_i \cdot \ln p_i 
\end{equation}
where, $N$ is the number of VM types, $p_i$ is the ratio of VM type $i$ of the total number of VMs. Then, the degree of heterogeneity is defined as follows:
\begin{equation}\label{eq:diversity_2}
      \eta = H / H_{max}
\end{equation}

Higher values of $\eta$ indicates a higher degree of heterogeneity in a cluster and vice versa. Once we know the degree of heterogeneity in a VM cluster, we can build the deallocation method accordingly. Algorithm~\ref{alg:deallocation} provides the pseudo-code proposed for the VM deallocation method. The method is triggered \emph{when} the deadline miss rate ($\gamma_{t}$) becomes less than the lower bound threshold $\alpha$, which is defined by the CVS2 user and represents how much the system can tolerate deadline miss rate in favor of cost-efficiency.

\begin{algorithm}[htb]
  \caption{Pseudo-code for the VM Deallocation Method}
  \label{alg:deallocation}
  \begin{algorithmic}[1]
    \INPUT\tikzmark{e} 
    \Statex $\alpha$: lower bound threshold for deadline miss rate

    \Statex
    \State $\gamma_t \leftarrow$ current deadline miss rate
    \If{$\gamma_{t} \leq \alpha$}
        \State calculate the utilization of each VM in the cluster
        \State find VM(s) with the lowest utilization
        \State resolve ties by choosing the least powerful VM(s) 
        \State $VM_j \leftarrow$ resolve ties by selecting the VM with the minimum remaining time to its charging cycle
        \State $\eta \leftarrow$ calculate the degree of heterogeneity
        \If{$\eta \geq \eta_{th}$ and $\rho_j \geq \rho_{th}$} 
            \State No deallocation
        \Else
            \State Deallocate $VM_j$
        \EndIf
    \EndIf     
  \end{algorithmic}
\end{algorithm}

The deallocation method is carried out in 4 main steps. In the first step, the VM(s) with the lowest utilization are chosen (lines 3 --- 4 in Algorithm~\ref{alg:deallocation}). In the second step, ties are broken by selecting the least powerful VM (line 5). If more than one VM remains, in the third step (line 6), ties are broken based on the VM with the minimum remaining time to its charging cycle. 

For a VM cluster that tends to a heterogeneous configuration (\ie $\eta \geq \eta_{th}$), the policy deallocates the selected VM (termed $VM_j$ in the algorithm) if its utilization is less than the VM utilization threshold (\ie $\rho_j < \rho_{th}$). The value of $\eta_{th}$ determines the boundary between homogeneous and heterogeneous configurations in a VM cluster. We experimentally realized that $\eta_{th}=0.4$ can discriminate homogeneous configurations from heterogeneous ones. The value of $\rho_{th}$ is determined by the CVS2 user based on its cost and performance trade-off. In contrast, in a VM cluster that tends to a homogeneous configuration, even if the utilization is high, the policy can deallocate $VM_j$ based on the deadline miss rate (lines 8 --- 12). 

It is worth noting that the deallocation method is also executed at the end of the charging cycle of the current VMs to deallocate VMs marked for deallocation. The reason for enacting VM termination at the end of the VM charging cycle is that the VM has already been paid for the whole charging cycle. Therefore, there is no benefit in terminating it before its charging cycle, even though it is recommended for deallocation. To implement this and to assure that no GOP task is left incomplete, the scheduler keeps track of each VM's remaining time to its charging cycle and the completion time of the tasks assigned to that VM. If a VM is marked for deallocation, before scheduler maps a new GOP task to it, the scheduler estimates the completion time of GOPs assigned to that VM, in addition to the completion time of the new GOP task. If the completion times are larger than the time remains to the VM's charging cycle, the GOP tasks are rescheduled on other VMs. Otherwise, the scheduler keeps sending GOP tasks to the VM, even though it is marked for deallocation.

%

\subsection{Remedial VM Provisioning Policy}
The periodic VM provision policy cannot cover request arrivals to the batch queue that occur in the interval of two provisioning events. 

To cope with the shortage of the periodic policy, we propose a lightweight remedial provisioning policy that can improve the overall performance of the VM Provisioner component. By injecting this policy into the intervals of the periodic provisioning policy, we can perform the periodic policy less frequently. 

In fact, the remedial provisioning policy provides a quick prediction of the system based on the state of the virtual queue. 
Recall that the Virtual Queue includes the distinction of streaming requests waiting for transcoding in the batch queue. Hence, the length of the Virtual Queue implies the intensity of streaming requests waiting for processing. 
Such long batch queue increases the chance of a QoS violation in the near future. Thus, our lightweight remedial policy only checks the size of the Virtual Queue (denoted \textit{$Q_s$}). Then, it uses Equation (\ref{eq:remedy}) to decide for the number of VMs that should be allocated.
\begin{equation}\label{eq:remedy}
n =  \lfloor\frac{\textit{$Q_s$}} {\theta \cdot \beta}\rfloor
\end{equation}

where $n$ is the \textit{number} of VM(s) that should be allocated; $Q_{s}$ is the size of the Virtual Queue. $\theta$ is a constant factor that determines the aggressiveness of the VM allocation in the policy. That is, lower values of $\theta$ leads to allocating more VMs and vice versa. In the implementation, we considered $\theta =10$. In the remedial policy, we allocate a VM type that, in general, provides a high performance per cost ratio (in the experiments, we used \texttt{c4.xlarge}).

Experiment results indicate that the remedial provisioning policy does not incur any extra cost to the stream service provider. Nonetheless, it increases the robustness of the QoS by reducing the average deadline miss rate and average startup delay (see Section~\ref{subsec:remedyexp}).
To verify the performance of the proposed methods, in the next section, we evaluate them in different configurations and under various workload conditions.


\section{Performance Evaluation}\label{sec:pe}
\subsection{Experimental Setup}\label{subsec:es}
We used CloudSim~\cite{pe_2}, a discrete event simulator, to model our system and evaluate performance of the scheduling methods and VM provisioning policies. To create a diversity of video streaming requests, we uniformly selected videos over the range of [10 , 600] seconds from a set of benchmark videos. We made the benchmarking videos publicly available for reproducibility purposes\footnote{The videos can be downloaded from: https://goo.gl/TE5iJ5}. We modeled our system based on the characteristics and cost of VM types in Amazon EC2. We considered \texttt{g2.2xlarge, c4.xlarge, r3.xlarge,} and \texttt{m4.large} in our experiments. The VMs represent the characteristics of various VM types offered by Amazon cloud and form a heterogeneous VM cluster.  

To simulate a realistic video transcoding scenario, using \texttt{FFmpeg}\footnote{https://ffmpeg.org}, we performed four different transcoding operations (namely codec conversion, resolution reduction, bit rate adjustment, and frame rate reduction) for each of the benchmarking videos. Then, the execution time of each transcoding operation was obtained by executing them on the different VM types.

To capture the randomness in the execution time of GOPs on cloud VMs, we transcoded each GOP 30 times and modeled the transcoding execution times of GOPs based on the Normal distribution\footnote{The generated workload traces are available publicly from: https://goo.gl/B6T5aj}.

To study the performance of the system comprehensively, we evaluated the system under various workload intensities. For that purpose, we varied the arrival rate of the video streaming requests from 100 to 1000 within the same period of time. The inter-arrival times of the requested videos are generated based on the Normal distribution, where the mean of inter-arrival time is based on the time divided by the number of requests and standard deviation is the mean divided by 3. All experiments of this section were run 30 times, and the mean and the 95\% of the confidence interval of the results are reported for each experiment. In all the experiments, we considered the values of $\alpha$ and $\beta$ equal to 0.05 and 0.1, respectively. That is, we consider that the SSP chose to keep the deadline miss rate between 5\% to 10\%. Any deadline miss beyond 10\% is considered as a \emph{QoS violation}. The QoS boundary is shown in the form of a horizontal line in the experiment results.

\subsection{Average Completion Time of Early GOP Tasks}
The goal of using utility-based mapping heuristics is to prioritize GOPs with high utility (\ie earlier GOPs in the stream) for reducing their completion time. Although this factor is extended in next experiments through evaluating the average startup delay. We conduct the experiment to further evaluate how this goal is satisfied when our utility-based scheduling methods with different mapping heuristics are applied. 

\begin{figure}[htbp]
\centering{
\includegraphics[width=0.75\columnwidth]{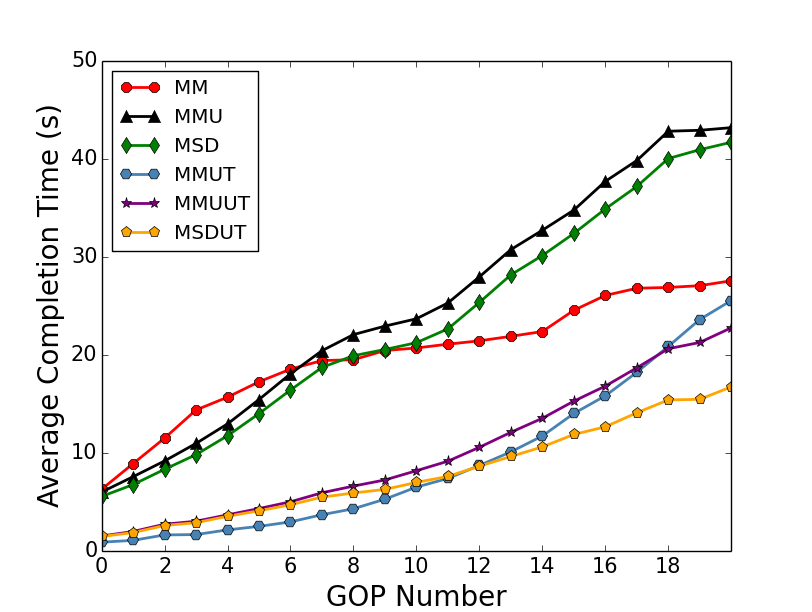}
\caption{Average completion time of early GOPs under different scheduling methods. The horizontal axis shows the GOP numbers in the video stream and the vertical axis shows the average completion time of GOPs. We used 1000 GOP tasks and the VM provisioning policies are applied.}
\label{fig:buffer}}
\end{figure}

In Figure~\ref{fig:buffer}, the horizontal axis is the GOP number of the first 20 GOPs in the benchmark video streams and the vertical axis is the average completion time of the GOPs in seconds. For this experiment, we have used 1000 GOP tasks and VM provisioning policies are in place.

Figure~\ref{fig:buffer} demonstrates that, in general, the utility-based heuristics provide a significantly lower average completion time. Among traditional heuristics, MM performs the best. This is because MM prioritizes the GOPs with short execution times, which results in faster processing in the system. We also observed that MSDUT performs better in compare with other utility-based heuristics, specifically for GOP numbers more than 15. This is because the dynamic VM provisioning policy works based on the tasks deadline miss rate. Since MSDUT favors tasks with short deadlines, many GOPs miss their deadlines as the system becomes busy. Therefore, it allocates more VMs that, in turn, reduces the average completion time of the GOPs.

\begin{figure*}[htbp]
\centering{
\subfigure[Startup delay under traditional heuristics]
{\includegraphics[width=0.31\textwidth]{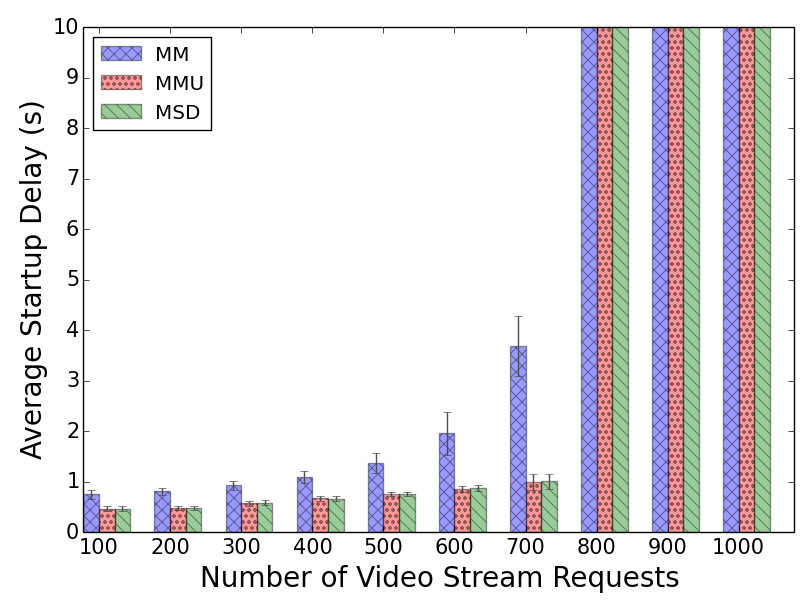}\label{fig:tss_stt}}
\subfigure[Deadline miss rate under traditional heuristics]
{\includegraphics[width=0.31\textwidth]{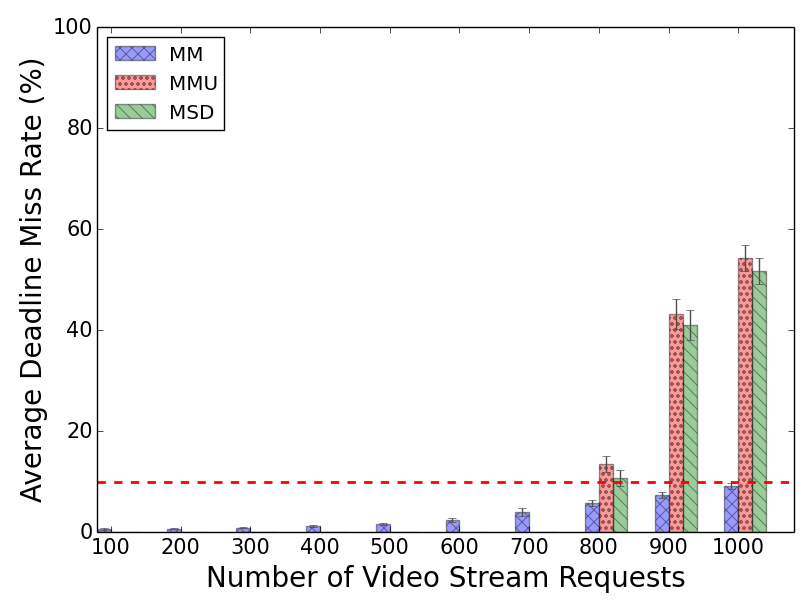}\label{fig:tss_dmr}}
\subfigure[Cost under traditional heuristics]
{\includegraphics[width=0.31\textwidth]{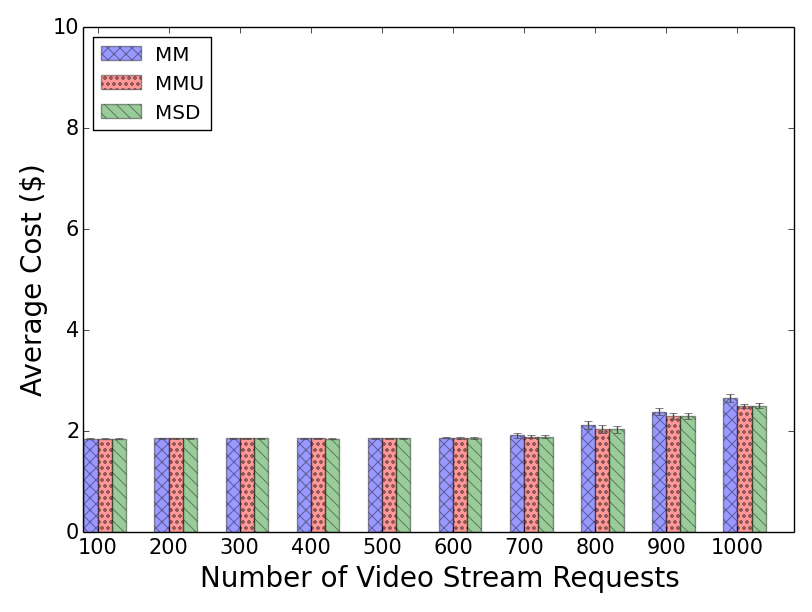}\label{fig:tss_cost}}
\subfigure[Startup delay under utility-based heuristics]
{\includegraphics[width=0.31\textwidth]{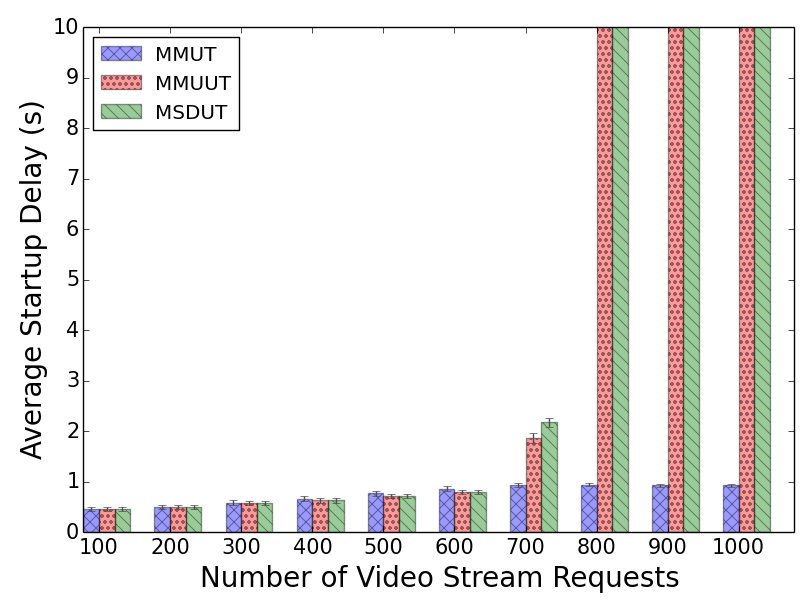}\label{fig:pss_stt}}
\subfigure[Deadline miss rate under utility-based heuristics]
{\includegraphics[width=0.31\textwidth]{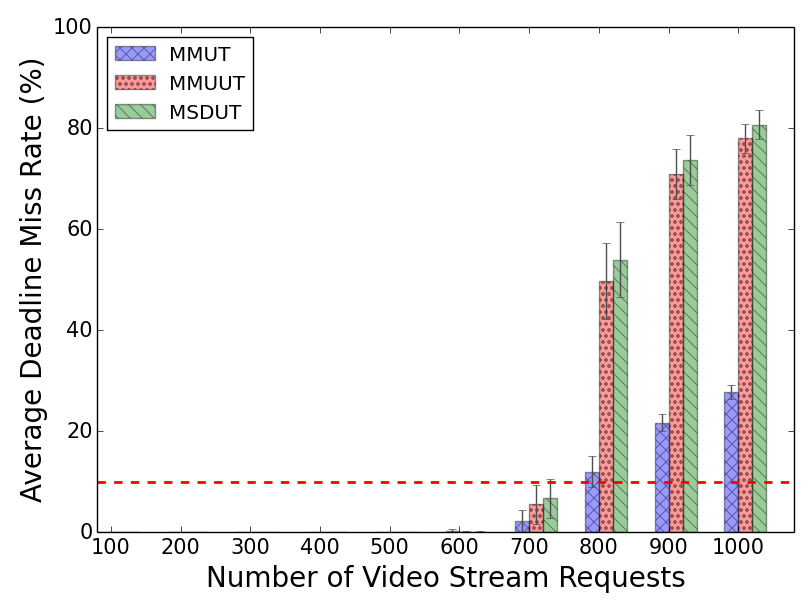}\label{fig:pss_dmr}}
\subfigure[Cost under utility-based heuristics]
{\includegraphics[width=0.31\textwidth]{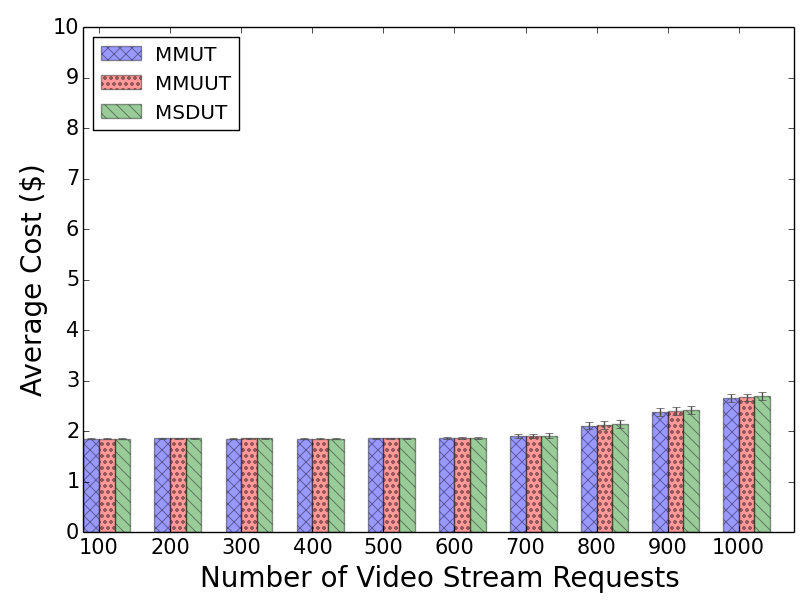}\label{fig:pss_cost}}
\caption{The results under utility-based mapping heuristics against those under traditional mapping heuristics when the number of video requests varies. Subfigures (a), (b), and (c), respectively, show the average startup delay, deadline miss rate, and the incurred cost under traditional mapping heuristics, while (d), (e), and (f) show the same factors under utility-based mapping heuristics are applied. The horizontal dashed line denotes the acceptable QoS boundary ($\beta$).}
\label{fig:ub_sche_static}}
\end{figure*}

\begin{figure*}[htbp]
\centering{
\subfigure[Startup delay under traditional heuristics]
{\includegraphics[width=0.31\textwidth]{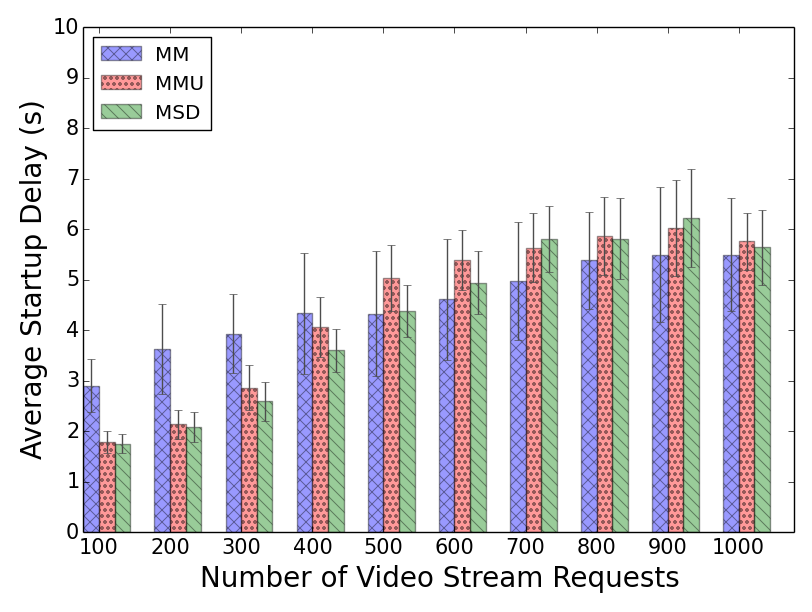}\label{fig:tsd_stt}}
\subfigure[Deadline miss rate under traditional heuristics]
{\includegraphics[width=0.31\textwidth]{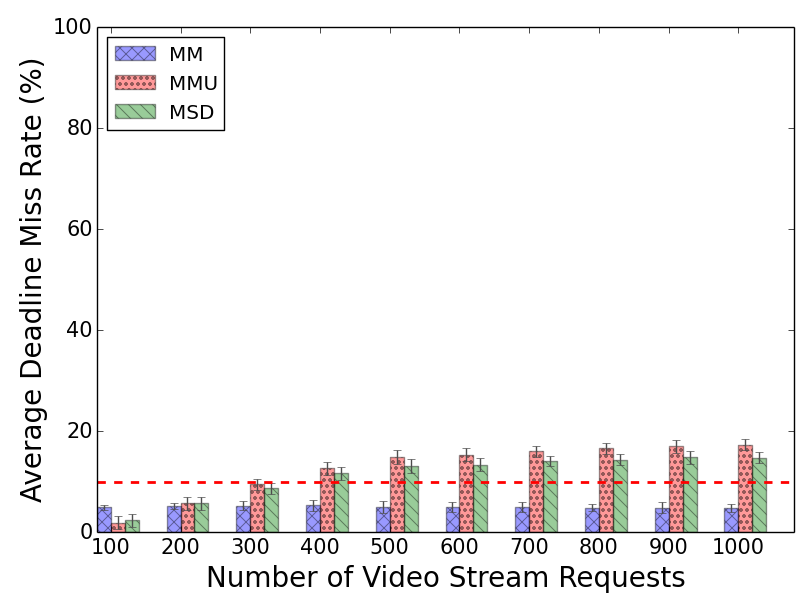}\label{fig:tsd_dmr}}
\subfigure[Cost under traditional heuristics]
{\includegraphics[width=0.31\textwidth]{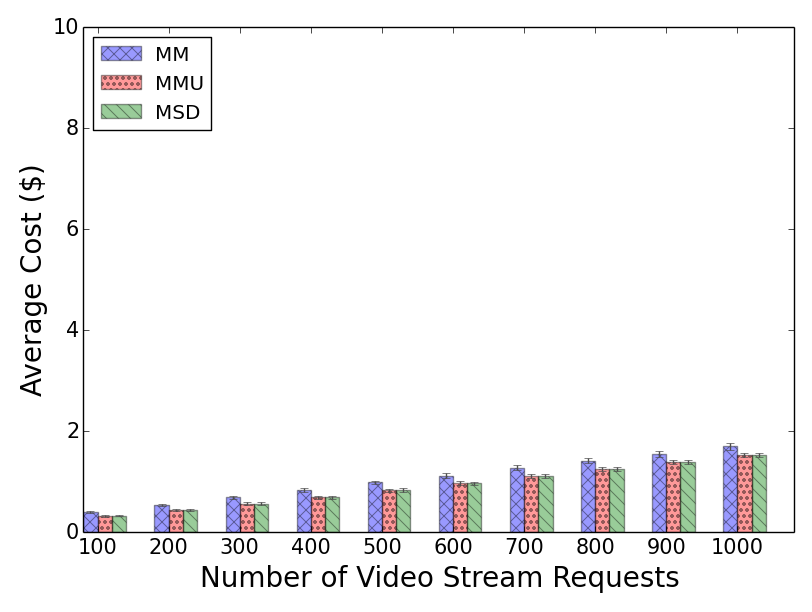}\label{fig:tsd_cost}}
\subfigure[Startup delay under utility-based heuristics]
{\includegraphics[width=0.31\textwidth]{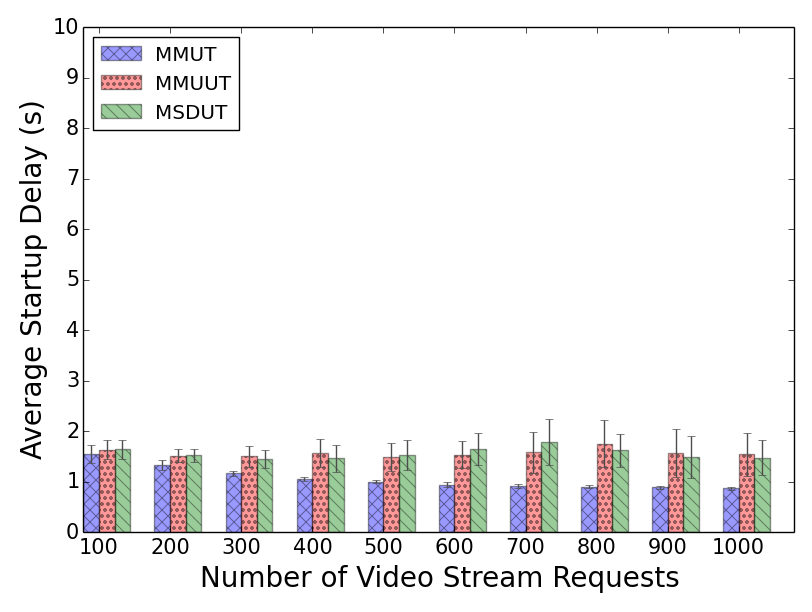}\label{fig:psd_stt}}
\subfigure[Deadline miss rate under utility-based heuristics]
{\includegraphics[width=0.31\textwidth]{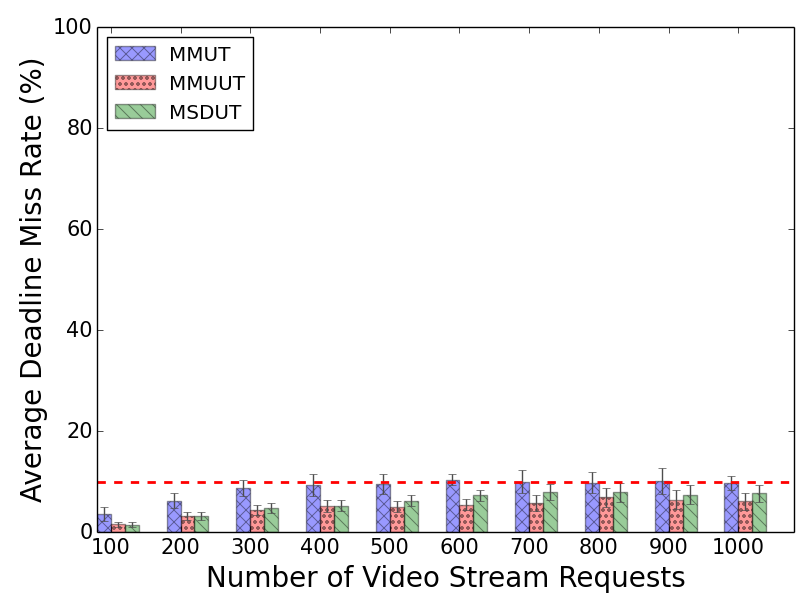}\label{fig:psd_dmr}}
\subfigure[Cost under utility-based heuristics]
{\includegraphics[width=0.31\textwidth]{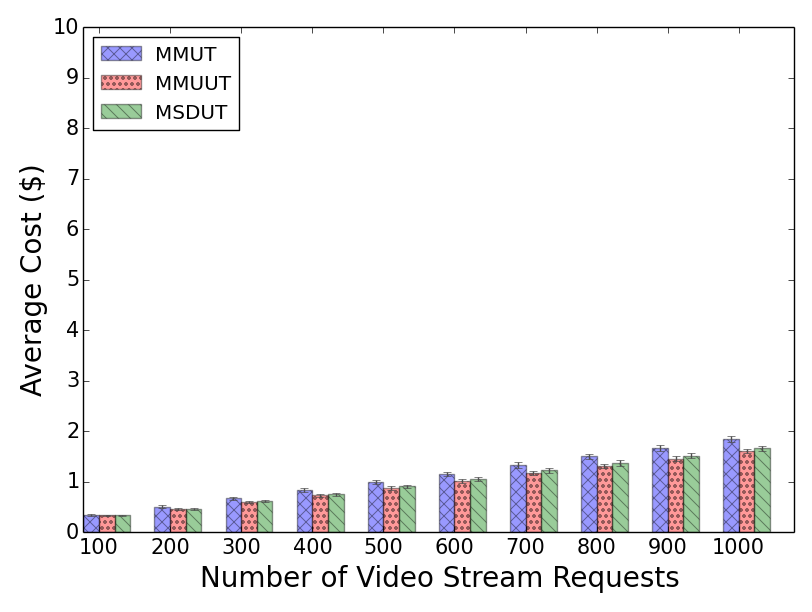}\label{fig:psd_cost}}
\caption{The results under utility-base mapping heuristics against those under traditional mapping heuristics when dynamic provisioning policies are applied. The X-axis indicates the number of streaming requests, and Subfigures (a), (b), and (c) show the average startup delay, deadline miss rate, and the incurred cost, respectively, under traditional mapping heuristics, while (d), (e), and (f) show the same factors under utility-based mapping heuristics. The horizontal dashed line indicates the acceptable QoS boundary ($\beta$).}
\label{fig:ub_sche_elastic}}
\end{figure*}

\subsection{Impact of Utility-based Mapping Heuristics}\label{subsec:utility_ex}
To evaluate the impact of utility-based mapping heuristics on QoS and cost, we compare them with the traditional mapping heuristics in two scenarios: (1) VM provisioning performed in the static way (Subsection~\ref{subsec:stat_het}) and (2) under the VM provisioning policies (Subsection~\ref{subsec:dyn_het}). To construct a static heterogeneous cluster, we allocate three VMs of each type. 

\subsubsection{Static Heterogeneous VM Cluster}\label{subsec:stat_het}
Figure~\ref{fig:ub_sche_static} compares the results of utility-based mapping heuristics with the traditional batch heuristics under a static heterogeneous VM cluster. For traditional mapping heuristics, Figures~\ref{fig:tss_stt} and~\ref{fig:tss_dmr} show that MM provides a significantly lower average deadline miss rate (by up to 40\%) than MSD and MMU, in particular when the system is more oversubscribed (\ie overloaded). However, MSD and MMU provide a lower average startup delay than MM. This is because both MSD and MMU function based on the deadline and the deadline of the startup GOPs is low since they are prioritized.

In Figure~\ref{fig:pss_dmr}, we observe that MMUT provides a significantly better average deadline miss rate (around 50\% when there are 1000 video requests) in comparison with MSDUT and MMUUT. More importantly, we can see, in Figure~\ref{fig:pss_stt}, that MMUT provides a low and stable startup delay in comparison with other heuristics even when the system is oversubscribed. This is because prioritizing shorter tasks in MMUT produces a lower average deadline miss rate which, in return, benefits the startup GOPs to be processed. 

We should note that although MMUT provides a lower start up delay, it yields a higher deadline miss rate than the traditional MM (see Figure~\ref{fig:ub_sche_elastic}). This is because the utility-based mapping heuristics prioritize GOPs with higher utility values (i.e., higher priority) to reduce the start up delay. This causes a higher deadline miss rate particularly when we use static resource allocation. As we will explain in the next section, utility-based mapping heuristics, in particular MMUT, significantly outperform traditional mapping heuristics, when accompanied with dynamic resource provisioning.

We do not observe any major cost difference for more intensive workloads. This is because in the static cluster, the workload can be handled within the same time period. When the system is oversubscribed, there is a minor increase in cost, as seen in Figure~\ref{fig:tss_cost} and Figure~\ref{fig:pss_cost}. This is because it takes a longer time to finish the processing of the tasks in those cases. 

\subsubsection{Dynamic Heterogeneous VM Cluster}\label{subsec:dyn_het}
Figure~\ref{fig:tsd_cost} demonstrates that, regardless of the mapping heuristic, the dynamic VM provisioning policy significantly reduces the incurred cost (up to ~80\% when the system is not oversubscribed) in comparison to the static heterogeneous VM cluster. The incurred cost increases as the VM provisioning policy needs to allocate additional VMs to maintain QoS robustness for more video streaming requests. 

In Figure~\ref{fig:tsd_stt}, we can observe that the average startup delay increases for traditional mapping heuristics. However, it is more stable in comparison with Figure~\ref{fig:tss_stt} with static heterogeneous VMs. This is because the VM provisioning policy adapts the VM provisioning to the workload intensity to meet the QoS demands of the stream viewers. 

Figures~\ref{fig:psd_stt}, ~\ref{fig:psd_dmr}, and ~\ref{fig:psd_cost} demonstrate the robustness resulted from applying the utility-based mapping heuristics together with the VM provisioning policies. That is, with the increase of the workload, the system all together produces a low and stable average startup delay and average deadline miss rate without incurring extra cost to the stream provider. In particular, we observe the average deadline miss rates of MMUUT and MSDUT have dramatically decreased. Normally, MMUUT and MSDUT lead to higher average deadline miss rates than MMUT. However, with the dynamic VM provisioning policies, the high deadline miss rates of MMUUT and MSDUT trigger the VM provisioning policies to allocate more VMs that, in turn, reduce the deadline miss rate. Nonetheless, the deadline miss rate of  MMUT is not sufficiently high enough to trigger the allocation method.

Further evaluations and comparisons against previous works are discussed in Appendix~\ref{apdx:exp}.
\subsubsection{Discussion}
We can summarize our findings about the proposed mapping heuristics (discussed in Subsections~\ref{subsec:stat_het} and~\ref{subsec:dyn_het}) as follows:
\begin{enumerate}
 \item In both static and dynamic heterogeneous VM provisioning: MMUT provides the lowest and the most stable average startup delay in compare with all other mapping heuristics.
 
 \item In both static and dynamic heterogeneous VM provisioning: The three proposed mapping heuristics incur approximately the same cost to the stream provider. 

 \item In static heterogeneous VM provisioning: MMUT results in a lower average deadline miss rate, in compare with MMUUT and MSDUT. 
 
 \item In dynamic heterogeneous VM provisioning: MMUUT and MSDUT outperform MMUT in terms of average deadline miss rate. Typically, MMUUT and MSDUT result in a higher deadline miss rate (as shown in Figure~\ref{fig:pss_dmr}, when a static VM provisioning is used). The reason for the opposite behavior of MMUUT and MSDUT, in dynamic VM provisioning, is that their higher deadline miss rate triggers allocating more VMs, hence, their deadline miss rate is decreased. It is worth noting that, although MMUT results in a higher deadline miss rate, it is still below the threshold provided by the video stream provider (see Figure~\ref{fig:psd_dmr}).
 
\end{enumerate}

\subsection{The Impact of VM Provisioning Policies}
\begin{figure*}[htbp]
\centering{
\subfigure[Comparison of average startup delay]
{\includegraphics[width=0.31\textwidth]{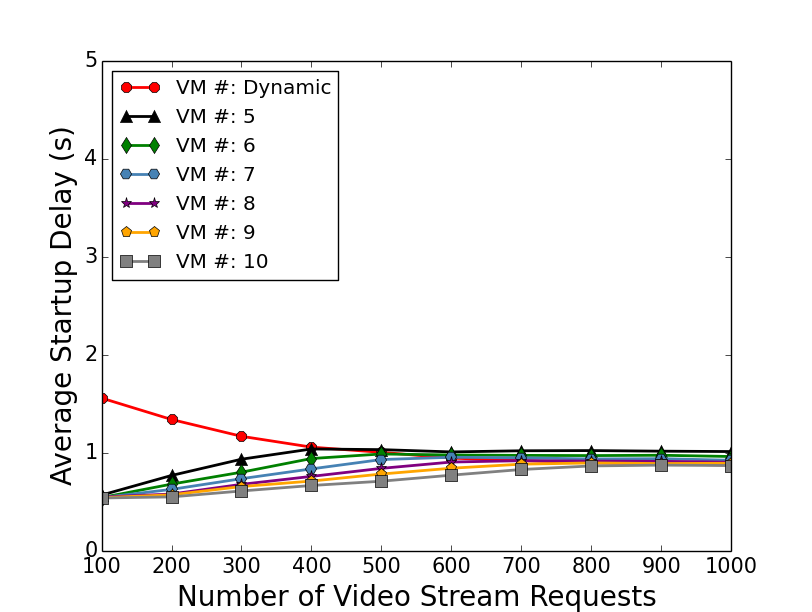}\label{fig:dvs_stt}}
\subfigure[Comparison of average deadline miss rate]
{\includegraphics[width=0.31\textwidth]{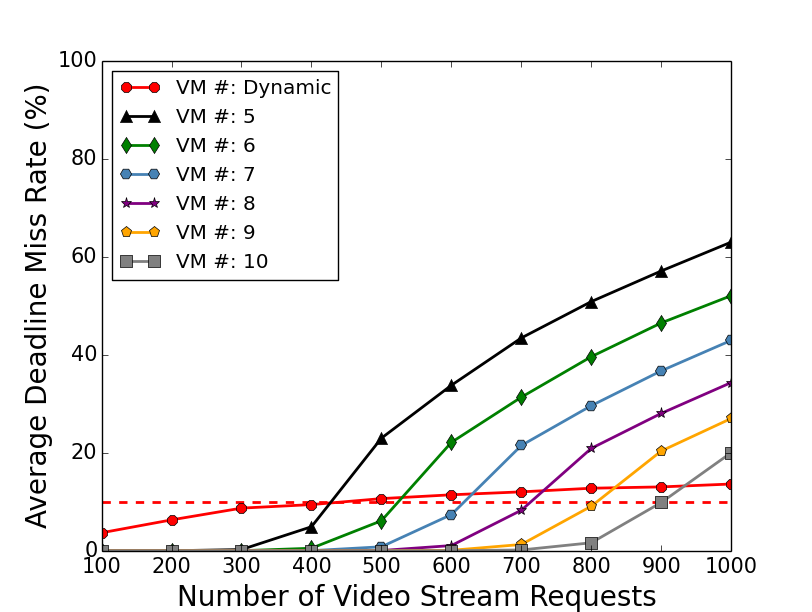}\label{fig:dvs_dmr}}
\subfigure[Comparison of average cost]
{\includegraphics[width=0.31\textwidth]{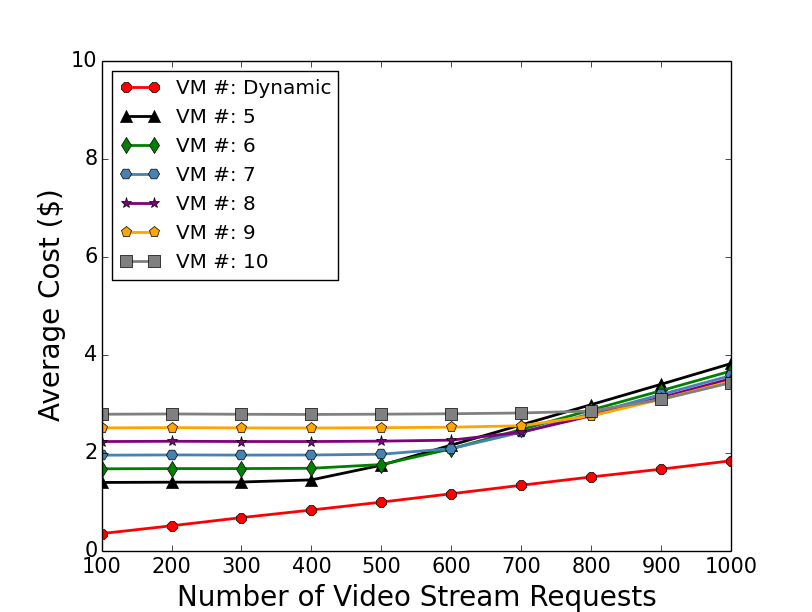}\label{fig:dvs_cost}}
\caption{Performance comparison under static and dynamic VM provisioning policies. Subfigure (a) illustrates the average startup delay, (b) shows the average deadline miss rate, and (c) demonstrates the incurred cost to the streaming provider under dynamic and static provisioning policies, with MMUT applied as the mapping heuristic.}
\label{fig:elastic_vs_static}}
\end{figure*}

To further investigate the performance of the proposed VM provisioning policies, we compare it against the case in which a static homogeneous VM cluster is deployed. For evaluation, we vary the number of streaming requests in the system from 100 to 1000.  In this experiment, we choose MMUT as the mapping heuristic. The reason for choosing MMUT is that, in general, it performs better than other heuristics both in static and dynamic VM provisioning. Albeit, MMUT does not outperform other heuristics in terms of deadline miss rate when dynamic VM provisioning is used (see Figure~\ref{fig:psd_dmr}). However, even in that case, it can still keep the deadline miss rate below the QoS threshold provided by the video stream provider For the static clusters, as it is shown in Figure~\ref{fig:elastic_vs_static}, we evaluate clusters with 5 to 10 VMs. In all of them we utilized GPU VM type (\texttt{g2.2xlarge}). We observed that the average startup delay, and the average deadline miss rate are too high when fewer VMs are allocated. Therefore, we do not include them in the graphs. We would like to note that we also used other VM types to compare against dynamic VM provisioning. However, their performances were even worse than the GPU type.

In Figure~\ref{fig:dvs_stt}, we can see that as the number of video requests increases, the average startup delay in all static policies grows. However, the dynamic VM provisioning policy provides a low and stable average startup delay. When the system is not oversubscribed (\ie number of stream requests less than 400), the dynamic provisioning policies provide a slightly higher startup delay than the static policy. The reason is that when the deadline miss rate is low, the VM provisioning policies allocate fewer VMs to reduce the incurred cost. Hence, new GOP tasks have to wait for transcoding. However, in the static policy, specifically with a large number of VMs, GOPs in the startup queue can be transcoded quickly, reducing the average startup delay.

Figure~\ref{fig:dvs_dmr} illustrates that the VM provisioning policies lead to low and stable average deadline miss rate in comparison with the static ones. In the static configuration, as the number of video requests increases, the average deadline miss rate grows dramatically. However, we notice that the average deadline miss rate with the dynamic VM provisioning policies remains stable, even when the system becomes oversubscribed. We can conclude from the experiment that the proposed VM Provisioner component in the \name enables the system to tolerate workload oversubscription. That is, it makes the system robust against the fluctuations in the arrival workload.

In addition to low and stable average startup delay and average deadline miss rate, Figure~\ref{fig:dvs_cost} shows that the dynamic VM provisioning policies reduce the incurred cost by up to 85\% when the system is not oversubscribed. Even when the system is oversubscribed (\ie with more than 500 streaming requests in the system) the dynamic VM provisioning policies reduced the cost to around 50\%. In fact, when the streaming request rate is low, VMs are under-utilized; however, in the static VM cluster, the streaming service provider still has to pay for them. In contrast, with the dynamic VM provisioning,  the system deallocates idle VMs when the deadline miss rate is low, which reduces the incurred cost significantly. As the number of streaming requests increases, more VMs of the appropriate types are created, and hence, the incurred cost of the dynamic VM provisioning policies approaches that of the static one. We can conclude that, from the cost perspective, our proposed VM provisioning policies are more efficient, particularly when the system is lightly loaded.

\subsection{Impact of the Remedial VM Provisioning Policy}\label{subsec:remedyexp}
\begin{figure}[htbp]
\centering{
\includegraphics[width=0.85\columnwidth]{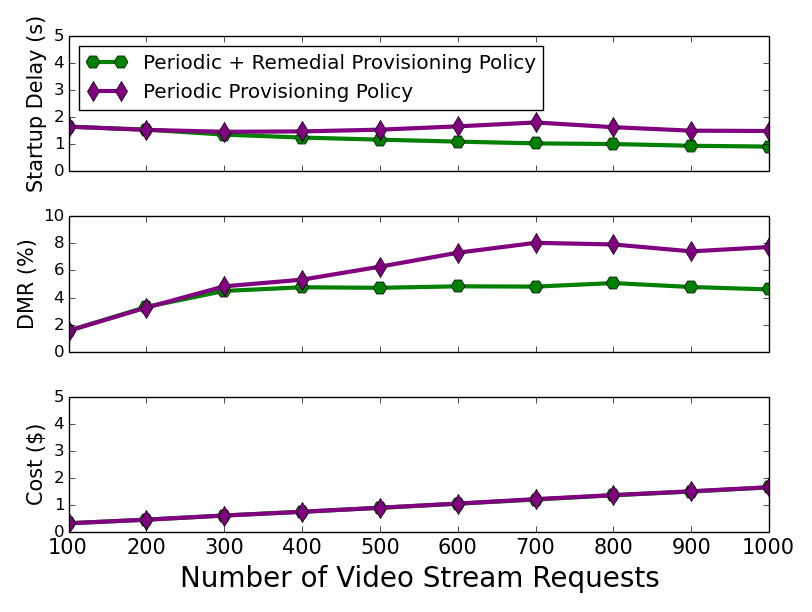}
\caption{Impact of the remedial VM provisioning policy on the startup delay, deadline miss rate (DMR) and the incurred cost.}
\label{fig:remedial}}
\end{figure}

To evaluate the efficacy of the remedial provisioning policy, we conduct an experiment on the dynamic VM provisioning policy in two scenarios: (A) when the VM Provisioner component uses both the periodic and remedial polices and (B) when only the periodic provisioning policy is in place. We measure QoS in terms of average Deadline Miss Rate (DMR), average startup delay, and the incurred cost when the number of streaming requests varies in the system (along the X-axis in Figure~\ref{fig:remedial}). In this experiment we assume that the MMUT mapping heuristic is utilized.
 
As illustrated in Figure~\ref{fig:remedial}, when the system is not oversubscribed (\ie fewer than 500 streaming requests), the difference between the two scenarios is negligible. This is because when streaming requests arrived between two provisioning events are not excessive, the VMs allocated by the periodic VM provisioning policy are sufficient to keep the QoS robust. 

Alternatively, when the system is oversubscribed, the number of streaming requests that arrive between two provisioning events is high and affects the prediction of the provisioning policy. Under this circumstance, as depicted in Figure~\ref{fig:remedial}, relying only on the periodic provisioning policy leads to a high deadline miss rate. Nonetheless, when the remedial VM provisioning policy is utilized even with the system is oversubscribed, the deadline miss rate remains stable. In addition, as it is shown in the last sub-figure of Figure~\ref{fig:remedial}, the remedial VM provisioning policy comes without incurring any extra cost to the stream provider.

\section{Related Work}\label{sec:rw}

\begin{figure}[htb] 
    \centering
    \includegraphics[width=3.5in]{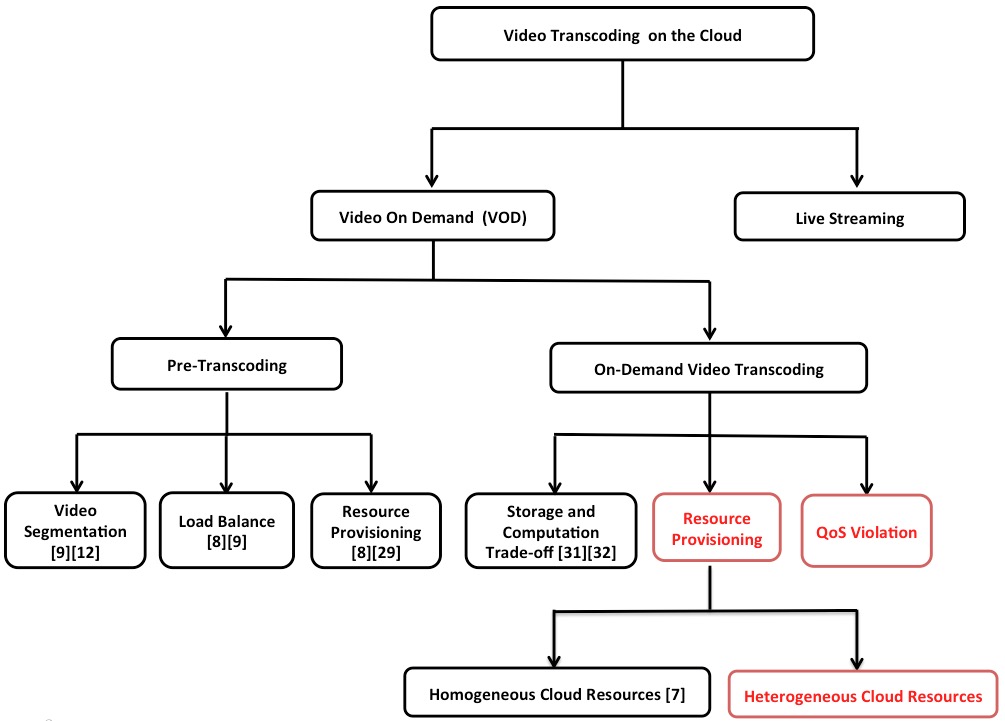}
    \caption{A taxonomy of video transcoding using cloud. Red blocks position the contributions of this work.}
    \label{fig:rw}
\end{figure}

Techniques, architectures, and challenges of video transcoding have been investigated by Ahmad \etal~\cite{intro_6} and Vetro \etal~\cite{intro_7}. 
Cloud-based video transcoding for VOD has been studied in~\cite{rw_10, rw_11}. However, they all investigated the case of offline transcoding (\ie pre-transcoding). 
A taxonomy of the researches undertaken on cloud-based video transcoding and the position of our contribution with respect to them is illustrated in Figure~\ref{fig:rw}.
\vspace{-0.3cm}
\vspace{-0.1cm}


Jokhio \etal~\cite{rw_13} present a computation and storage trade-off strategy for cost-efficient video transcoding in the cloud. The trade-off is based on the computation cost versus the storage cost of the video streams. They determine how long a video should be stored or how frequently it should be re-transcoded from a given source video. Zhao~\etal~\cite{rw_18} take the popularity, computation cost, and storage cost of each version of a video stream into account to determine versions of a video stream that should be stored or transcoded. The earlier studies demonstrate that it is possible to transcode infrequently accessed videos streams in an on-demand manner~\cite{manish14}. However, they do not explore the possible ways to carry out the on-demand transcoding efficiently by utilizing appropriate scheduling methods and VM provisioning policies.



In systems with dynamical task arrival, task scheduling can be performed either in an \emph{Immediate} or a \emph{Batch} mode~\cite{rw_9}. In the former, the tasks are mapped to processing machines as soon as they arrive to the scheduler,  whereas in the latter, few tasks are collected in a batch queue and are scheduled at the same time. Amini Salehi \etal~\cite{rw_17} have compared these scheduling types in heterogeneous computing systems and concluded that the batch-mode significantly outperforms the immediate-mode. The reason is that, in the batch-mode, tasks can be shuffled and they do not have to be assigned in the order they arrived. Accordingly, we consider batch-mode mapping in the scheduling component of the \name architecture. 
It is noteworthy that the current batch-mode scheduling heuristics (\eg see those in~\cite{rw_9}) and even those in the immediate-mode cannot fulfill the QoS requirements of on-demand video transcoding applications, mainly in terms of the startup delay. 

To consider the startup delay, in~\cite{pre_3}, a startup queue was considered to prioritize the first few GOPs in video streams. Alternatively, in this paper, we improve the startup queue model by assigning a utility value to each GOP. To minimize the startup delay, the earlier GOPs in a video stream are assigned higher utility values. 


Ashraf \etal~\cite{rw_3} propose a stream-based admission control and scheduling approach using a two-step prediction model to foresee the upcoming streams' rejection rate through predicting the waiting time at each machine. Later, a job scheduling method is utilized to drop some video segments to prevent video transcoding jitters. However, they do not consider minimizing the startup delay of video stream using a heterogeneous cluster of VMs.


Previous works on cloud-based VM provisioning for video transcoding (\eg~\cite{rw_11, rw_12}) mostly consider the case of off-line transcoding. Thus, their focuses are mainly on reducing makespan (\ie total transcoding times) and the incurred costs.

Netflix adopts the \emph{scale up early, scale down slowly} principle for its VM provisioning~\cite{rw_14} on Amazon EC2. It periodically checks the utilization of its allocated VMs. The allocated VMs are scaled up by 10\%, if their utilization is greater than 60\% for 5 minutes. They are also scaled down by 10\%, if the VMs utilizations is less than 30\% for 20 minutes. Lorido \etal~\cite{rw_14} categorize current auto-scaling techniques into five main families: static threshold-based rules, control theory, reinforcement learning, queuing theory, and time series analysis. Then, they utilize the classification to carry out a literature review of proposals for auto-scaling in the cloud.



In our earlier works~\cite{pre_3, pre_4}, a QoS-aware VM provisioning policy was proposed for on-demand video transcoding.Nonetheless, the policy did not consider heterogeneous types of VMs offered by cloud providers. They just consider one type of VM (\ie a homogeneous cluster of VMs) and try to minimize the incurred cost to the stream provider. Given the affinity between different transcoding types and VM types, VM provisioning policies are required to allocate and deallocate from heterogeneous VM types to minimize the incurred cost. This will enable the creation of a dynamically-formed VM cluster that changes its configurations based on the arriving transcoding requests. The current work is different from~\cite{pre_3} in several other ways too. We provide a method to quantify heterogeneity of a VM cluster and use it in deallocation policy of the VM cluster. We provide a method to quantify the suitability of each VM type for various transcoding operations. We develop new scheduling heuristics that are QoS-aware and are tailored for heterogeneous computing systems. We also provide a utility function that prioritizes GOPs in a video stream based on their position in the stream.
\section{Conclusions and Future Work}\label{sec:conclusion}
In this paper, we proposed the \name streaming engine for on-demand video transcoding. In particular, we developed the Task Scheduler and VM Provisioner components of \name. The components are aware of the viewers' QoS demands and aim to maintain QoS robustness while minimizing the incurred cost to the SSP. The components take advantage of the heterogeneous VMs, offered by the cloud providers with diverse prices. The Scheduler minimizes the startup delay and the deadline violations of the streams. The VM Provisioner is cost-aware in allocating/deallocating heterogeneous VMs. 
Experiment results demonstrate that proposed scheduling reduces the average startup delay and the deadline miss rate. In addition, heterogeneous VM provisioning reduces the incurred cost by up to 85\%, particularly, when the system is not oversubscribed. The VM provisioning is robust against uncertainties in the arrival of streaming requests, without incurring any extra cost to the provider. 

The \name architecture is useful for SSPs to utilize cloud services and offer on-demand transcoding of video streams with a low cost. In future, we will extend the admission control  to be failure-aware. We will also consider multiple cloud scenarios for faster video delivery.

\section*{Acknowledgments}
This research was supported by the Louisiana Board of Regents under grant number LEQSF(2016-19)-RD-A-25. This is a substantially extended version of a paper presented at the IEEE/ACM International on Conference on Cluster, Cloud and Grid Computing (CCGrid '16)~\cite{pre_3}.

%
%
%
%
%

%

\balance
\bibliographystyle{IEEEtran}
\bibliography{IEEEabrv,reference}

\vspace{-0.6cm}
\begin{IEEEbiography}[{\includegraphics[width=1in,height=1.25in,clip,keepaspectratio]{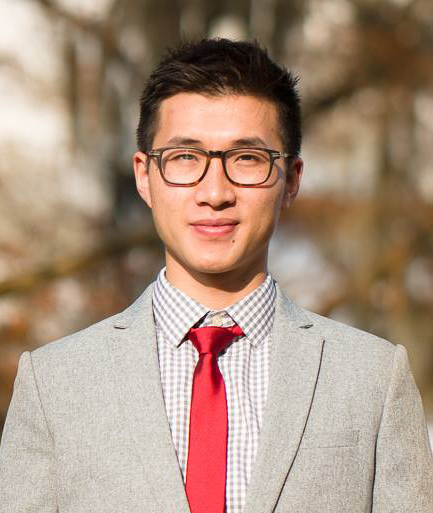}}]%
{Xiangbo Li}
received his Ph.D. degree in computer engineering from University of Louisiana at Lafayette in 2016. He is currently working as video engineer at Brightcove Inc., a cloud based online video platform company. He is an expert in cloud-based video encoding, transcoding, and packaging.
\end{IEEEbiography}
\vspace{-0.6cm}
\begin{IEEEbiography}[{\includegraphics[width=1in,height=1.25in,clip,keepaspectratio]{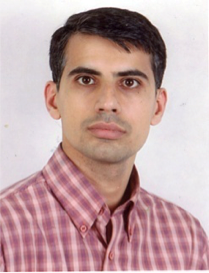}}]%
{Mohsen Amini Salehi}
received his Ph.D. in Computing and Information Systems from Melbourne University, Australia, in 2012. He is currently an Assistant Professor and director of the High Performance Cloud Computing (\href{http://hpcclab.org/}{HPCC}) laboratory, School of Computing and Informatics at University of Louisiana Lafayette, USA. His research focus is on Distributed and Cloud computing including heterogeneity, virtualization, resource allocation, and security. 
\end{IEEEbiography}
\vspace{-0.6cm}
\begin{IEEEbiography}[{\includegraphics[width=1in,height=1.25in,clip,keepaspectratio]{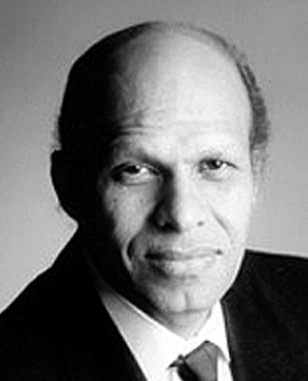}}]%
{Magdy Bayoumi}
received the BSc and MSc degrees in electrical engineering from Cairo University, Egypt, the MSc degree in computer engineering from Washington University, St. Louis, and the Ph.D. degree in electrical engineering from the University of Windsor, Ontario. He was the Vice President for Conferences of the IEEE Circuits and Systems (CAS) Society. He is the recipient of the 2009 IEEE Circuits and Systems Meritorious Service Award and the IEEE Circuits and Systems Society 2003 Education Award. 
\end{IEEEbiography}
\vspace{-0.6cm}
\begin{IEEEbiography}[{\includegraphics[width=1in,height=1.25in,clip,keepaspectratio]{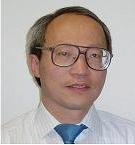}}]%
{Nain-Feng Tzeng} (M86-SM92-F10) received the Ph.D. degree in Computer Science from the University of Illinois at Urbana-Champaign. Since 1987, he has been with Center for Advanced Computer Studies, University of Louisiana at Lafayette, where he is currently a professor. He was on the editorial boards of the IEEE Transactions on Parallel and Distributed Systems, 1998 --- 2001, and IEEE Transactions on Computers, 1994 --- 1998. 
\end{IEEEbiography}
\vspace{-0.6cm}
\begin{IEEEbiography}[{\includegraphics[width=1in,height=1.25in,clip,keepaspectratio]{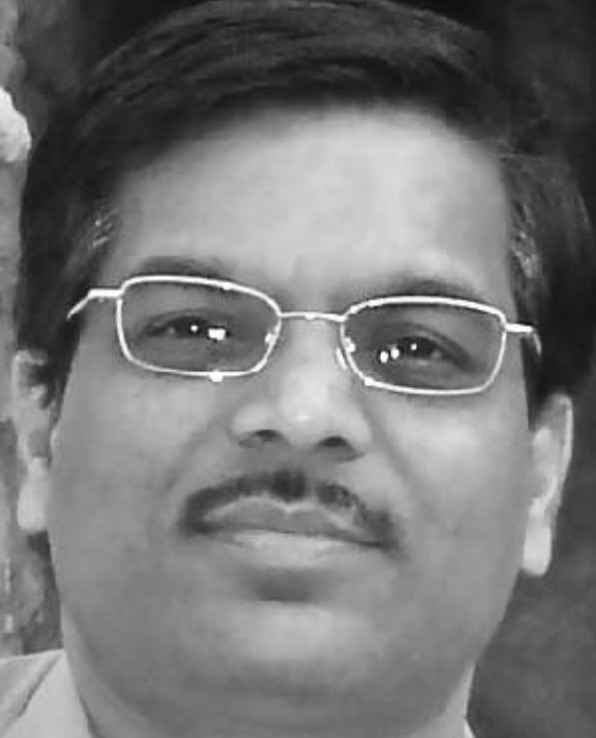}}]%
{Rajkumar Buyya} is a Fellow of IEEE, Professor of Computer Science and Software Engineering, Future Fellow of the Australian Research Council, and Director of the Cloud Computing and Distributed Systems (CLOUDS) Laboratory, School of Computing and Information Systems, at the University of Melbourne, Australia. He is one of the highly cited authors in computer science and software engineering worldwide. Microsoft Academic Search Index ranked Dr. Buyya as \#1 author in the world (2005-2016) for both field rating and citations evaluations in the area of Distributed and Parallel Computing.
\end{IEEEbiography}


\end{document}